\documentclass[12pt, preprint]{emulateapj}
\usepackage{amsmath}
\usepackage{lscape}

\shortauthors{Shi et al.}

\begin{document}

\title{Extended Schmidt Law: Role Of Existing Stars In Current Star Formation}

\author{Yong  Shi\altaffilmark{1},  George Helou\altaffilmark{1},  Lin
Yan\altaffilmark{1},      Lee      Armus\altaffilmark{1},      Yanling
Wu\altaffilmark{1},     Casey     Papovich\altaffilmark{2},    Sabrina
Stierwalt\altaffilmark{1}  } 
\altaffiltext{1}{Infrared  Processing and
Analysis   Center,   California    Institute   of   Technology,   1200
E. California Blvd, Pasadena, CA 91125} 
\altaffiltext{2}{George P. and
Cynthia   Woods  Mitchell  Institute   for  Fundamental   Physics  and
Astronomy, Department of Physics and Astronomy, Texas A\&M University,
College Station, TX 77843-4242}

\begin{abstract}

We propose an ``extended Schmidt law'' with explicit dependence of the
star formation efficiency (SFE=SFR/$M_{\rm  gas}$) on the stellar mass
surface density ($\Sigma_{\rm star}$).   This relation has a power-law
index of 0.48$\pm$0.04  and an 1-$\sigma$ observed scatter  on the SFE
of 0.4  dex, which  holds over  5 orders of  magnitude in  the stellar
density  for  individual   global  galaxies  including  various  types
especially  the  low-surface-brightness  (LSB) galaxies  that  deviate
significantly  from the  Kennicutt-Schmidt  law. When  applying it  to
regions at sub-kpc  resolution of a sample of  12 spiral galaxies, the
extended Schmidt  law not  only holds for  LSB regions but  also shows
significantly  smaller  scatters   both  within  and  across  galaxies
compared  to  the  Kennicutt-Schmidt  law.   We argue  that  this  new
relation points to  the role of existing stars  in regulating the SFE,
thus  encoding better  the  star formation  physics.  Comparison  with
physical  models of  star formation  recipes shows  that  the extended
Schmidt law can  be reproduced by some models  including gas free-fall
in  a  stellar-gravitational  potential  and  pressure-supported  star
formation.  By  implementing this new  law into the analytic  model of
gas accretion  in $\Lambda$  CDM, we show  that it can  re-produce the
observed main  sequence of  star-forming galaxies (a  relation between
the SFR and stellar mass) from $z$=0 up to $z$=2.

\end{abstract}

\keywords{galaxies: evolution -- galaxies: starburst -- stars: formation -- ISM: molecules -- ISM: HI }

\section{Introduction}

Stars  form from the  cold interstellar  medium (ISM).   The resulting
stellar mass  growth, chemical enrichment  and energy feedback  to the
ISM  and  intergalactic  medium  (IGM)  are key  processes  of  galaxy
formation and evolution.  Understanding how  stars form is thus one of
the central  questions in galactic and  extragalactic astronomy.  Star
formation  invokes   a  series  of  complicated   processes  from  gas
accretion, gas  cooling and H$_{2}$  formation to the  final molecular
cloud collapse  and stellar feedback.  Empirical  scaling laws between
star  formation   and  gas   reservoirs  provide  critical   tests  of
our modeling of  the  above  various  processes  and  have  crucial
applications  to studies  of  galaxy formation  and  evolution in  the
cosmological context.

In  the  pioneering work  of  \citet{Schmidt59},  a  simple power  law
relation is proposed  to relate the star formation  rate (SFR) density
to the gas density:
\begin{equation}
 \rho_{\rm SFR} = A\rho_{\rm gas}^{N_{\rm SFR}}
\end{equation}
where $\rho_{\rm SFR}$ is the  SFR volume density, $\rho_{\rm gas}$ is
the  volume density  of  total  cold gas  including  HI and  molecular
hydrogen (H$_{2}$) and  $N_{\rm SFR}$ is the power  index while $A$ is
simply  assumed  to  be  constant.  \citet{Kennicutt89,  Kennicutt98a}
demonstrated  this  unambiguously  in  its  observable  form  (surface
density)  with  61 nearby  spiral  galaxies  and 36  infrared-selected
nuclear  starburst  regions that  span  a  large  dynamic range  and
concluded:
\begin{equation}
 \Sigma_{\rm SFR} \propto \Sigma_{\rm gas}^{1.4\pm0.15}
\end{equation}
which is often  referred as the Kennicutt-Schmidt (KS)  law.  It makes
general sense that the gas  reservoir determines how
many stars can form, i.e., the  gas density plays the dominant role in
regulating SFR. With this basic relation between SFR and gas, the star
formation efficiency (SFE=SFR/gas-mass in this study) follows:
\begin{equation}
 {\rm SFE}=\frac{\Sigma_{\rm SFR}}{\Sigma_{\rm gas}} \propto \Sigma_{\rm gas}^{N_{\rm SFE}}\text{, with } N_{\rm SFE}=0.4
\end{equation} 
It should be noted  that the KS law does not hold  for the whole range
of  gas  densities.   An  accompanying  rule  is  the  star  formation
threshold,   introduced  to  explain   the  fact   that  the   SFR  is
significantly lower than predicted by  the KS law at low gas densities
($<$$\sim$1-10 M$_{\odot}$ pc$^{-2}$), e.g., in low-surface-brightness
(LSB)   galaxies   or   regions   far   outside   the   optical   disk
\citep[e.g.][]{Martin01, Wyder09, Bigiel08, Roychowdhury09}.  Other forms of the
Schmidt law  \footnote{In this study  we expand the definition  of the
Schmidt law  to include  any relationship that  invokes SFRs  and gas,
which shares the  initial idea that Schmidt proposed.}  have also been
proposed,   such  as   those   that  invoke   the  dynamical   factors
\citep{Silk97, Elmegreen97, Boissier03}:
\begin{equation}
 {\rm SFE} \propto \frac{1}{\tau_{\rm dyn}}
\end{equation}
where $\tau_{\rm  dyn}$ is the
orbital dynamical  timescale.  This relation has  been demonstrated to
predict the  SFR as well  as the KS  law when considering  the orbital
timescale \citep{Kennicutt98a},  while the break  may still show  up at
the low density end \citep{Wyder09}.

Although  the above two  Schmidt relations  are valid  for a  range of
galaxy types  both in the  local universe \citep[e.g.][]{Kennicutt98a,
Bigiel08}  and  at  high-z  \citep[e.g.][]{Daddi10b, Genzel10},  it  is
surprising that  they only invoke  the gas component while  in reality
various  ISM  and  stellar   components  are  intimately  involved  in
processes of  converting gas into  new stars. Many of  these processes
are  related to  existing  stars  that form  over  the whole  galaxy's
history.  For  example, the gravitational effects of  stellar bars can
remove  gas  angular momentum  and  increase  nuclear  SFRs in  normal
galaxies  \citep[e.g.][]{Sersic67,  Ho97}.   Their  effects  are  also
recognized  in numerical  simulations of  gas-rich galaxy  merging and
thought  to be  the  main factor  to  determine how  much  gas can  be
converted   into  stars   \citep{Hopkins09a,  Hopkins09b}.    The  gas
hydrostatic pressure produced by  gas self-gravity and stellar gravity
is further shown empirically to  be related to the H$_{2}$-to-HI ratio
\citep{Elmegreen94,  Wong02, Blitz04,  Blitz06}.  Besides  the stellar
gravity, the metal outputs of  stellar evolution are the main coolants
of  gas  and thus  star  formation  should  show dependencies  on  the
metallicity.   Theoretical works  have  indicated significantly  lower
SFEs  at low  metallicity \citep{Krumholz09,  Gnedin10}.   Dust grains
that form from metals catalyze  the  H$_{2}$  formation and shield
it from radiation destruction \citep[for    a    review,
see][]{Hollenbach99}.  Current proposed  Schmidt relations do not have
any hint for these effects, as they only invoke the gas component.

To evaluate the importance of  existing stars in the empirical scaling
law, we demonstrate the existence of a tight relationship between SFEs
and stellar  mass surface densities, referred as  the extended Schmidt
law.  This relation  not only predicts the SFE and SFR  as well as the
KS law for galaxies and spatially-resolved regions ($\sim$1 Kpc sizes)
where the  KS law works, but  also holds for LSB  galaxies and regions
where the  KS law fails.   Similar close links between  star formation
and total  stars have  been recognized in  previous works  for certain
galaxy  types or  limited stellar  mass surface  density  ranges.  For
example, \citet{Hunter98}  have shown for  radial azimuthally-averaged
quantities in LSB  galaxies that the stellar mass  density is the only
quantity  spatially  related  to   the  SFR  density.   Similar  close
associations between  stellar masses and  SFR densities are  also seen
within  and  among   galaxies  by  \citet{Ryder94},  \citet{Brosch98},
\citet{Hunter04} and  recently by \citet{Leroy08}  for specific galaxy
types or  limited density ranges.  Here, we  demonstrate this intimate
association directly by  showing a tight relation between  the SFE and
stellar   mass    density   over   a   large    dynamic   range.    In
\S~\ref{sample_data},  we  present   the  sample  selection  and  data
collection.    We   show    the   result   in   \S~\ref{result}.    In
\S~\ref{discussion}, we first compare  the extended Schmidt law to the
model of  the star formation  recipe and then discuss  its implication
for  the  main sequence  of  star-forming  galaxies.  Conclusions  are
presented  in \S~\ref{paper_conclusion}.   Throughout  this paper,  we
assume  $H_{0}$=70   km  s$^{-1}$  Mpc$^{-1}$,   $\Omega_{0}$=0.3  and
$\Omega_{\Lambda}$=0.7.

\section{Sample and Data}\label{sample_data}

\subsection{Individual Global Galaxies}\label{sample_ind_gal}

Our whole  sample is listed in  Table~\ref{tab_SAMP_COHI} and composed
of  five   sub-samples  including  low-redshift   late-type  galaxies,
early-type  galaxies,  LSB galaxies,  luminous  infrared galaxies  and
high-z objects.  The CO and HI data are collected from the literature.
Molecular gas  masses are derived from  the CO by  assuming a constant
CO-to-H$_{2}$  conversion factor of  $X_{\rm CO}$=2.0$\times$10$^{20}$
cm$^{-2}$ (K  km s$^{-1}$)$^{-1}$ or 3.17 M$_{\odot}$  pc$^{-2}$ (K km
s$^{-1}$)$^{-1}$   or   7845$d_{\rm   L}^{2}$   M$_{\odot}$   (Jy   km
s$^{-1}$)$^{-1}$ where  $d_{\rm L}$ is the luminosity  distance in Mpc
\citep{Dickman86, Solomon87, Tacconi08}.  Different conversion factors
for    mergers    and    non-mergers    are    also    discussed    in
\S~\ref{res_ind_gal}.  A factor of 1.36 is further included to account
for  the presence  of  heavier elements  in  both the  H$_{2}$ and  HI
masses.  The  SFR data are collected  from the same  references as the
gas  data (see  Table~\ref{tab_SAMP_COHI}), and  all are  corrected to
Chabrier    initial    mass    function    (IMF)    where    SFR$_{\rm
Chabrier}$$\approx$SFR$_{\rm     Kroupa}$=0.66SFR$_{\rm     Salpeter}$
\citep{BC03}.

The stellar masses are measured by fitting the \citet{BC03} population
synthesis  model to  the multi-band  SED with  Chabrier  IMF following
\citet{Shi08}.  The  details of the  parameters to produce  the models
are  listed  in   Table~\ref{BC_model}.   The  model  of  \citet{BC03}
generally produces consistent color and mass-to-light ratio at various
bands compared  to others  \citep{Vazquez05}. However, it  still lacks
accurate  evolutionary tracks  of thermally  pulsing  asymptotic giant
branch (AGB)  stars.  At  the stellar age  around 0.1-3 Gyr  where AGB
stars are  prominent, the  mass-to-light ratio in  the near-IR  can be
overestimated by $\sim$60\%  \citep{Maraston06, Bruzual07}. The effect
of  this on  the  result  of this  paper  is shown  to  be small  (see
\S~\ref{res_ind_gal}).   To  minimize systematic  errors  that may  be
caused  by different template  population models,  the range  of input
parameter spaces, numerical method and  etc, we have applied the above
method  to   all  of   our  objects  except   for  23   galaxies  from
\citet{Leroy08} whose masses are derived from IRAC 3.6 $\mu$m emission
in that  work.  For these objects,  the published gas  and SFR surface
densities are  defined within 1.5$R_{25}$ where  the available optical
photometry does  not exist.   For the spiral  objects in  their sample
where the majority  ($>$95\%) of the optical light  is enclosed within
$R_{25}$ \citep{Courteau96},  the median offset  between their stellar
masses and ours  by fitting models to optical/IRAC  photometry is only
0.1  dex.  The broad-band  wavelengths used  to calculate  the stellar
masses for all objects are listed in Table~\ref{tab_SAMP_COHI} and the
majority covers both the optical and near-IR bands.

The  surface densities  of gas,  SFR and  stellar masses  are measured
within   the   same   aperture   and   corrected   for   inclinations.
Table~\ref{tab_SAMP_COHI}  lists the  aperture  definitions and  their
relations to  optical isophotal radii  ($R_{\rm 25}$) if  available in
\citet{Paturel03}\footnote{http://leda.univ-lyon1.fr/}.       Different
apertures are  adopted for different  galaxy types, partly  because of
the heterogeneous nature  of the sample but also  because of different
light structures  of individual galaxy types.  Since  a star formation
law  describes  how star  forms  from gas,  an  ideal  aperture for  a
galaxy-averaged star formation relationship should enclose the majority of
star formation  or gas. This  certainly results in  different aperture
sizes for  galaxies with different types, e.g.,  compact apertures for
ultra  luminous  infrared galaxies  (ULIRGs)  and  wide  ones for  low
surface-brightness     (LSB)     galaxies.      As     discussed     in
\S~\ref{res_ind_gal}, the  extended Schmidt law depends  little on how
exactly an aperture is defined.   We estimated a typical error of 0.30
dex for each  quantity (SFR, gas and stellar  mass densities) based on
our  own experiences  of measurements,  while local  LIRGs  and high-z
galaxies have higher uncertainties  ($\sim$0.5 dex) due to low spatial
resolution.

1.  Late-type galaxies: This subsample includes 61 and 18 objects from
\citet{Kennicutt98a} and \citet{Leroy08}, respectively. For 11 objects
included in  both samples, the  data from the  latter is used  in this
work    because    of   higher    data    quality.    Galaxies    from
\citet{Kennicutt98a} have published SFR and gas surface densities that
are   defined    to   be   within   the    optical   isophotal   radii
\citep[$R_{25}$,][]{deVaucouleurs76}.   The   SFR  is  based   on  the
extinction-corrected  H$\alpha$  emission  with $A_{H\alpha}$=1.1  mag
(see  their paper  for the  equation). The  optical/near-IR photometry
used  to derive  stellar masses  are  collected from  NED.  Since  the
aperture of $R_{25}$ encloses  the majority ($>$95\%) of optical light
for spiral  galaxies \citep{Courteau96}, the  aperture-matched stellar
mass  density  is  thus  defined  by  dividing  the  total  mass  with
${\pi}R_{25}^{2}$.   All  spiral  galaxies from  \citet{Leroy08}  have
published  gas, SFR  and stellar  mass densities,  where  the aperture
radius  is defined to  be 1.5$r_{25}$.  The SFR  is measured  from the
combination of  FUV and  24 $\mu$m emission  (see their paper  for the
equation). For 11 objects from  this study also included in the sample
of \citet{Kennicutt98a},  we checked the median offset  in the stellar
mass between ours and theirs is only 0.1 dex.

2.  Early-type galaxies: Low level star formation has been detected in
circumnuclear  regions of  many early-type  galaxies \citep{Shapiro10,
Wei10, Crocker11}.  Although the difficulty of SFR measurements due to
contamination from  large populations of  old stars and  possible AGN,
these objects are  shown to follow more or  less the Kennicutt-Schmidt
law.  We here included nine  objects from \citet{Wei10} and 10 objects
from \citet{Crocker11}.  \citet{Wei10} have published aperture-matched
H$_{2}$ mass from CO data and  SFR from a combination of 24 $\mu$m and
UV emission  following \citet{Leroy08}.  For  six of nine  objects, we
estimated the HI mass from the  available VLA HI map in the literature
(see  Table~\ref{tab_SAMP_COHI}) and  found that  the HI  mass  in the
CO-aperture is $<$  10\% of the total HI. Based  on this, we therefore
included three more  objects with the total HI mass  $<$ 1.6 times the
H$_{2}$  mass, which implies  the HI  mass within  the CO  aperture is
smaller  than  16\% of  H$_{2}$.   Objects  in \citet{Crocker11}  have
available  aperture-matched total  gas mass  and SFR  measurements. As
recommended  in   that  work,  we  have   used  24$\mu$m+H$\alpha$  or
PAH+H$\alpha$ if  available and radio+H$\alpha$ for  two more objects,
where they  have used  formula calibrated in  \citet{Kennicutt09} (see
their  Table 4).  For  objects from  both works,  the aperture-matched
stellar mass density is estimated  in this work by fitting the stellar
synthesis models to  the aperture-matched 2MASS J, H  and K photometry
and SDSS optical photometry for most of them.

3. Low-surface-brightness  (LSB) galaxies: We defined  the LSB objects
as  an independent  subsample  as  they show  deviations  from the  KS
law. The  subsample includes all  the 19 objects  from \citet{Wyder09}
and 5  gas-rich dwarf galaxies from \citet{Leroy08}  that have stellar
mass surface  densities below 1 M$_{\odot}$ pc$^{-2}$  that is roughly
the  upper  limit  of  the  \citet{Wyder09}  sample.   \citet{Wyder09}
measured  the surface densities  of the  SFR and  gas mass  through UV
emission (using  \citet{Kennicutt98b} equation) and  HI, respectively,
where the aperture  is defined to be the minimum  of the maximum radii
of  UV and  HI emission.   For  these objects,  the contribution  from
obscured  star  formation  should  be  negligible  to  the  total  SFR
\citep[e.g.][]{Hinz07},  while the  HI should  dominate the  total gas
mass \citep[e.g.][]{Matthews05, Leroy08}.  We derived the stellar mass
from the SDSS photometry  while additional near-IR photometry from NED
was  also  used   for  two  objects.   As  the   published  SDSS  data
under-estimate the sizes and thus brightness of these objects, we have
re-measured  five-band  SDSS photometry  in  the  aperture adopted  in
\citet{Wyder09}.  The majority of  these objects have optical emission
out to the adopted radius as observed by SDSS (see the $r$-band radial
profile in \citet{Wyder09}).  The  comparison of our r-band photometry
to those  measured by \citet{Wyder09} shows  $<$10\% discrepancy.  All
dwarf  galaxies  from  \citet{Leroy08}  have published  gas,  SFR  and
stellar mass densities within 1.5$r_{25}$.

4.  Local LIRGs (z=0 LIRGs):  We have collected seven local LIRGs from
the  literature  with  two  criteria: (1)  the  spatially-resolved  CO
interferometer  images  are available  and  well  resolved (i.e.   the
deconvolved   size   is  larger   than   the   resolution);  (2)   the
interferometer  fluxes recover the  majority ($>$90\%)  of single-dish
measurements. As listed in Table 1, the aperture to define the surface
density is the  maximum extent of the CO emission.   For the total gas
density, we  have neglected  the contribution from  the atomic  gas in
these gas rich objects.  Since star formation takes place in molecular
clouds, we  assume all star formation  is included in  the CO aperture
where   the   SFR   is   based   on   the   IR   luminosity   assuming
\citet{Kennicutt98b}  relations  and corrected  to  Chabrier IMF.   To
measure the aperture-matched stellar mass density, we first calculated
the total mass by fitting stellar models to the UV/optical/near-IR SED
collected from NED, and then measured the part in the CO aperture with
the HST/ACS-F814w image assuming  a constant mass-to-light ratio, which
can cause additional stellar mass uncertainty ($\sim$0.3 dex).

5.   High-redshift  star-forming galaxies  (high-z  SFGs) and  merging
sub-millimeter   galaxies   (high-z  SMGs):   We   here  included   21
optically/near-IR selected  star-forming galaxies (EGS,  BzK and BXMD)
and 7  sub-millimeter objects from  \citet{Genzel10}. The SFR  and gas
are available  in that work. The SFR  for the EGS is  estimated from a
combination  of  extinction-corrected  H$\alpha$/[O  II]/GALEX-UV  and
Spitzer  24 $\mu$m  emission.  For  the BzK,  it is  a  combination of
extinction-corrected UV and 24  $\mu$m emission. For the BXMD objects,
it is  from extinction-corrected H$\alpha$  emission while the  SFR of
the  SMG is  from 850  $\mu$m emission  (see \citet{Genzel10}  for the
equation)  .   We  measured  stellar  masses  by  our  own  to  reduce
systematic errors  among different studies.  The median  offset of our
stellar masses  compared to  the literature data  are -0.16  dex, 0.34
dex, -0.55  dex and 0.00  dex for EGS,  BzK, BXMD and  submm galaxies,
respectively  (see  \citet{Genzel10}, \citet{Daddi10a},  \citet{Erb06}
and  \citet{Hainline10}, respectively).   The large  offsets  for BXMD
($z\sim$2.2) and  BzK ($z\sim$1.5) are mainly  due to the  lack of the
rest-frame  near-IR photometry,  which could  cause  large differences
when different template  SEDs are used. With the  unified stellar mass
measurement, the scatter  among them in the extended  Schmidt law does
become smaller.  The  galaxy size is defined as  the half light radius
obtained from the fit to  H$\alpha$, optical/UV or CO images.  Similar
to \citet{Genzel10},  all densities are defined within  the half light
radius   ($R_{1/2}$),   e.g.,   $\Sigma_{\rm   star}$   =   0.5M$_{\rm
star}$/(${\pi}R^{2}_{1/2}$).   Here   we  did  not   account  for  the
difference  in the  half-light radius  among  SFR, gas  and stars  for
individual galaxies, since these three half light radii are on average
quite close to each other \citep[e.g.][]{Swinbank10}.

\subsection{Individual Regions In Spiral Galaxies}\label{data_ind_reg}

With the advent of high spatial-resolution SFR, gas and stellar images
of nearby galaxies, the star formation law at sub-kpc scales have been
studied  extensively  \citep[e.g.][]{Wong02,  Jogee05,  Crosthwaite07,
Schuster07, Kennicutt07, Bigiel08, Leroy08}. The general conclusion is
that the relationship  between SFR and total gas  varies strongly both
within galaxies  and across different  objects. This implies  that the
physics other  than those directly  related to the total  gas strongly
affect the SFE of the total gas. To test the idea of the SFE regulated
by existing  stars as  proposed by the  extended Schmidt law,  we have
carried measurements of SFR, gas and  stellar masses in a sample of 12
spiral  galaxies as  listed in  Table~\ref{samp_ind_region}.  They are
derived from  The HI Nearby  Galaxy Survey \citep[THINGS,][]{Walter08}
and The SIRTF Nearby Galaxies Survey \citep[SINGS,][]{Kennicutt03}.

For each object,  the SFR and gas mass  are measured within individual
750$\times$750   pc$^{2}$  regions  across   the  main   optical  disk
(semi-major axis $<$ $R_{25}$), with the technical procedure basically
following  \citet{Bigiel08} \citep[also see][]{Leroy08}  but corrected
to  our  IMF  and  CO-to-H$_{2}$  conversion  factors.   The  SFR  is
estimated  from  combination of  GALEX  far-UV \citep{GildePaz07}  and
Spitzer 24 $\mu$m  \citep{Kennicutt03}, with the 3-$\sigma$ lower-limit
around  10$^{-4}$ M$_{\odot}$/yr/kpc$^{2}$.  The  gas mass  is derived
from a combination of THINGS HI data \citep{Walter08} and BIMA SONG CO
J=1-0 map  \citep{Helfer03}, with the limiting  surface density around
1.5  M$_{\odot}$ pc$^{-2}$. Since  the spatial  coverage of  BIMA SONG
does  not  extend significantly  beyond  the HI-to-H$_{2}$  transition
radius  where the  total  gas is  dominated  by HI  emission, we  have
extrapolated the CO  data to pixels without observations  based on the
observed CO/HI  ratio as  a function of  semi-major axis.   The result
changes  little if  adopting the  mean CO/HI  ratio as  a  function of
radius   derived  from  \citet{Leroy08}.   For  seven   galaxies  with
significantly extended CO coverage in \citet{Bigiel08}, our derived slope
of the KS law is consistent with theirs within 20\%.  The stellar mass
is  estimated  based  on   the  mass-to-light  ratio  at  Spitzer  3.6
$\mu$m. The SINGS 3.6 $\mu$m  image is further subtracted by the median
sky level after binning  to the resolution of 750$\times$750 pc$^{2}$,
while the  3-$\sigma$ sky fluctuation gives the  limiting stellar mass
surface density around 1  M$_{\odot}$ pc$^{-2}$.  To reduce the effect
of radial gradient  of stellar age, extinction and  metallicity on the
stellar mass measurement, the 3.6$\mu$m mass-to-light ratio is derived
based  on  the optical  color,  for  which  the theoretical  trend  is
computed from  our stellar population synthesis models  with solar and
0.25  solar   metallicity.   For   five  galaxies  with
available  SDSS images, the  optical color  is defined  to be  the g-r
color, while the mean trend of these objects is used for the remaining
objects.

\section{RESULTS}\label{result}

\subsection{The Extended Schmidt Law For Individual Galaxies}\label{res_ind_gal}

\begin{figure*}
\epsscale{0.9}
\plotone{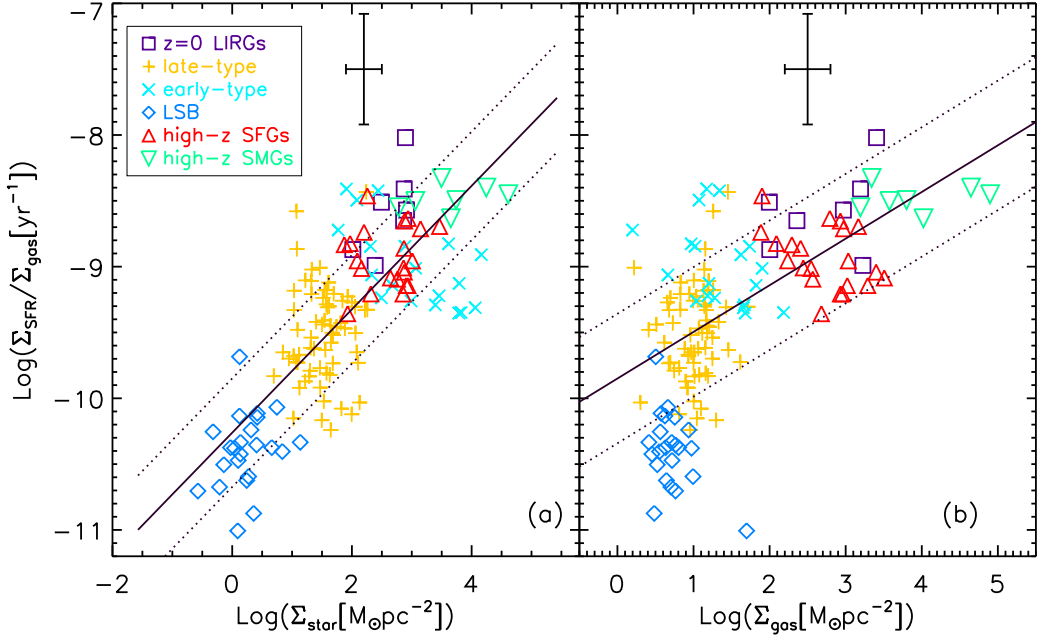}
\caption{\label{SFE_SBMstar} {\bf  (a):} The star-formation efficiency
(=$\Sigma_{\rm SFR}$/$\Sigma_{\rm gas}$) as  a function of the stellar
mass surface  density.  {\bf (b):}  The SFE as  a function of  the gas
density.  The  solid and dotted  lines are the intrinsic  best-fit and
observed 1-$\sigma$      scatter,     respectively,      as      listed     in
Table~\ref{fit_result}. The  fit to the  SFE-$\Sigma_{\rm gas}$ (right
panel) is done excluding LSB and early-type galaxies. Typical error bars are plotted.}
\end{figure*}

\begin{figure*}
\epsscale{0.9}
\plotone{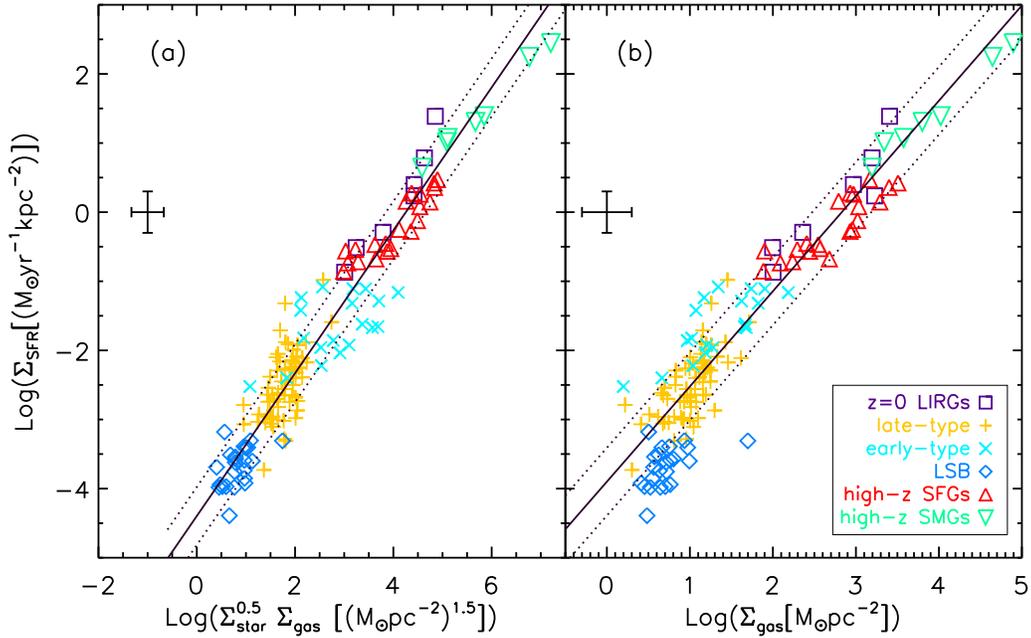}
\caption{\label{SFR_MgasMstar} Comparison between the extended Schmidt law
 and Kennicutt-Schmidt relationship  in the ability
to  predict the SFR.   The solid  and dotted  lines are  the intrinsic
best-fit   and  observed 1-$\sigma$  scatter,   respectively,  as   listed  in
Table~\ref{fit_result}. The fit to  the KS law (right panel) is done
excluding LSB and early-type  galaxies. Typical error bars are plotted.}
\end{figure*}

By  treating   $\Sigma_{\rm  SFR}$   as  a  dependent   variable,  and
$\Sigma_{\rm gas}$  and $\Sigma_{\rm star}$  as independent variables,
we search  for the best  power-law exponents relating them.   A linear
regression fit (IDL regress.pro) gives:
\begin{equation}{\label{eq_fit_three}}
 \Sigma_{\rm   SFR} \propto \Sigma_{\rm gas}^{1.13\pm0.05}\Sigma_{\rm star}^{0.36\pm0.04}
\end{equation}  
The most  important result of  the fit is  that the derived  index for
$\Sigma_{\rm star}$ is not zero that  would be expected by the KS law. 
Secondly,  the derived  exponent for $\Sigma_{\rm  gas}$,  namely  
approximately  unity, suggests  a  clear physical   implication   of  the   relation,   i.e.,   that  the   SFE
(=$\Sigma_{\rm SFR}/\Sigma_{\rm  gas}$) is  related to  the stellar
mass surface density.  We thus  carry out directly the fit between SFE
and  $\Sigma_{\rm  star}$   as  shown  in  Figure~\ref{SFE_SBMstar}(a)
through a Bayesian approach  to linear regression \citep{Kelly07} that
also  accounts  for uncertainties  in  both  variables.   As shown  in
Table~\ref{fit_result}, the best-fit gives:
\begin{equation}{\label{eq_sfe_star}}
 \frac{\rm SFE}{\rm yr^{-1}} = 10^{-10.28\pm0.08}(\frac{\Sigma_{\rm star}}{\rm M_{\odot}pc^{-2}})^{0.48\pm0.04}
\end{equation} 
where errors of  the best-fit parameters are the  intrinsic ones.  The
SFE-$\Sigma_{\rm star}$  relation is  obviously different from  the KS
law,  although both  of them  could be  used to  predict the  SFE.  By
invoking  the  stellar   mass  density,  the  SFE-$\Sigma_{\rm  star}$
correlation describes another scaling relation for star formation with
emphasis on the role of  the existing stellar component in the current
star formation activity. In what  follows, we refer to the relation in
Equation~\ref{eq_sfe_star} or the  SFE-$\Sigma_{\rm star}$ relation as
the extended Schmidt law  as it includes additional parameter (stellar
surface density).

\begin{figure}
\epsscale{1.2}
\plotone{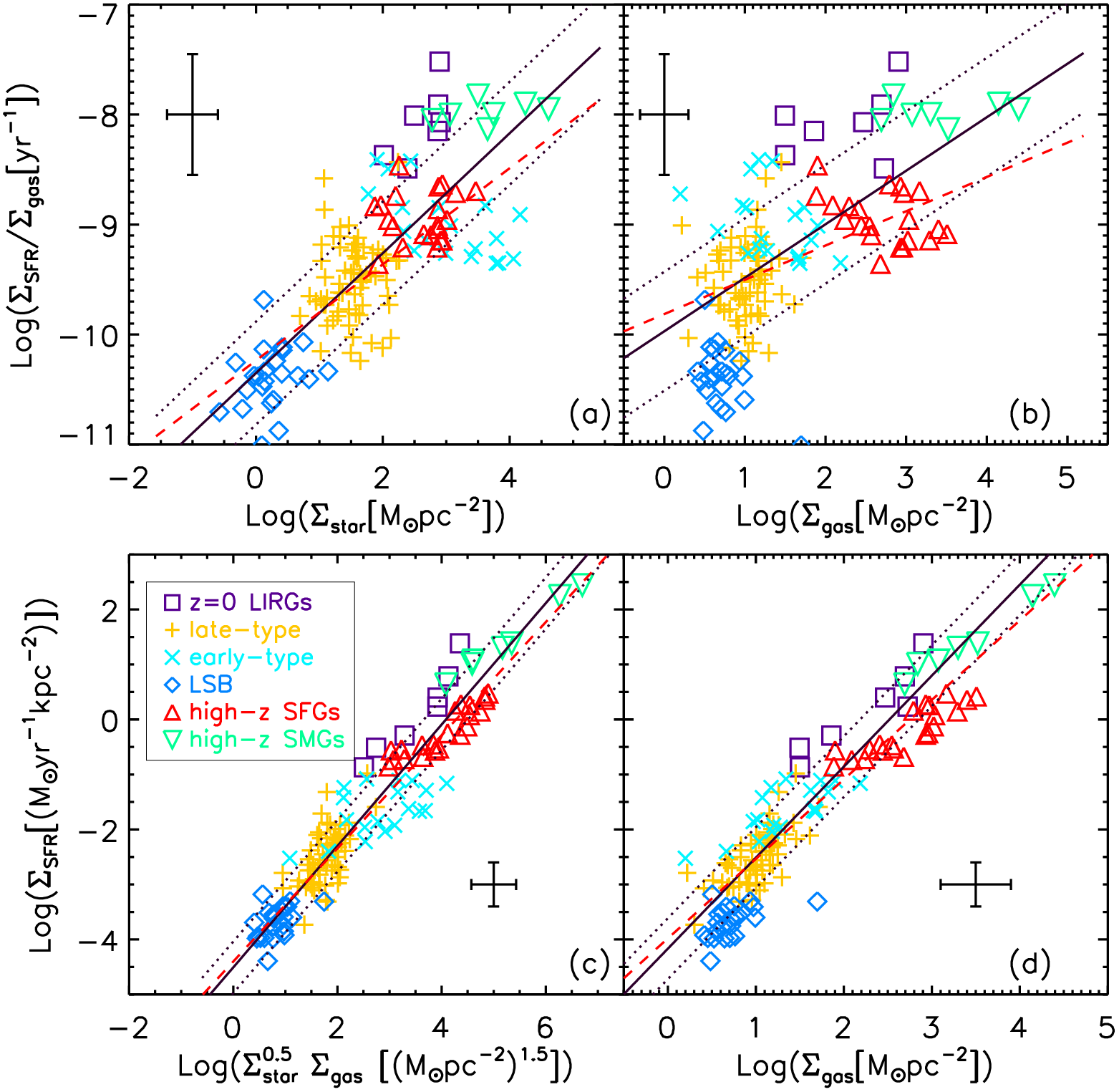}
\caption{\label{result_diff_a} Similar to Figure~\ref{SFE_SBMstar} and
Figure~\ref{SFR_MgasMstar}    but   using    different   CO-to-H$_{2}$
conversion  factors  for  normal galaxies  ($\alpha$=3.17  M$_{\odot}$
pc$^{-2}$ (K km s$^{-1}$)$^{-1}$) and merging objects (local LIRGs and
high-z sub-millimeter galaxies;  $\alpha$=1.0 M$_{\odot}$ pc$^{-2}$ (K
km s$^{-1}$)$^{-1}$).   The solid and  dotted lines are  the intrinsic
best-fit and observed 1-$\sigma$ scatter, respectively, to all data points but
excluding LSB and early-type  galaxies for panel (b)  and (d). The  dashed lines show
the best fit  to all the non-mergers but excluding  LSB  and early-type  ones for panel
(b) and (d).  }
\end{figure}

\begin{figure*}
\epsscale{0.9}
\plotone{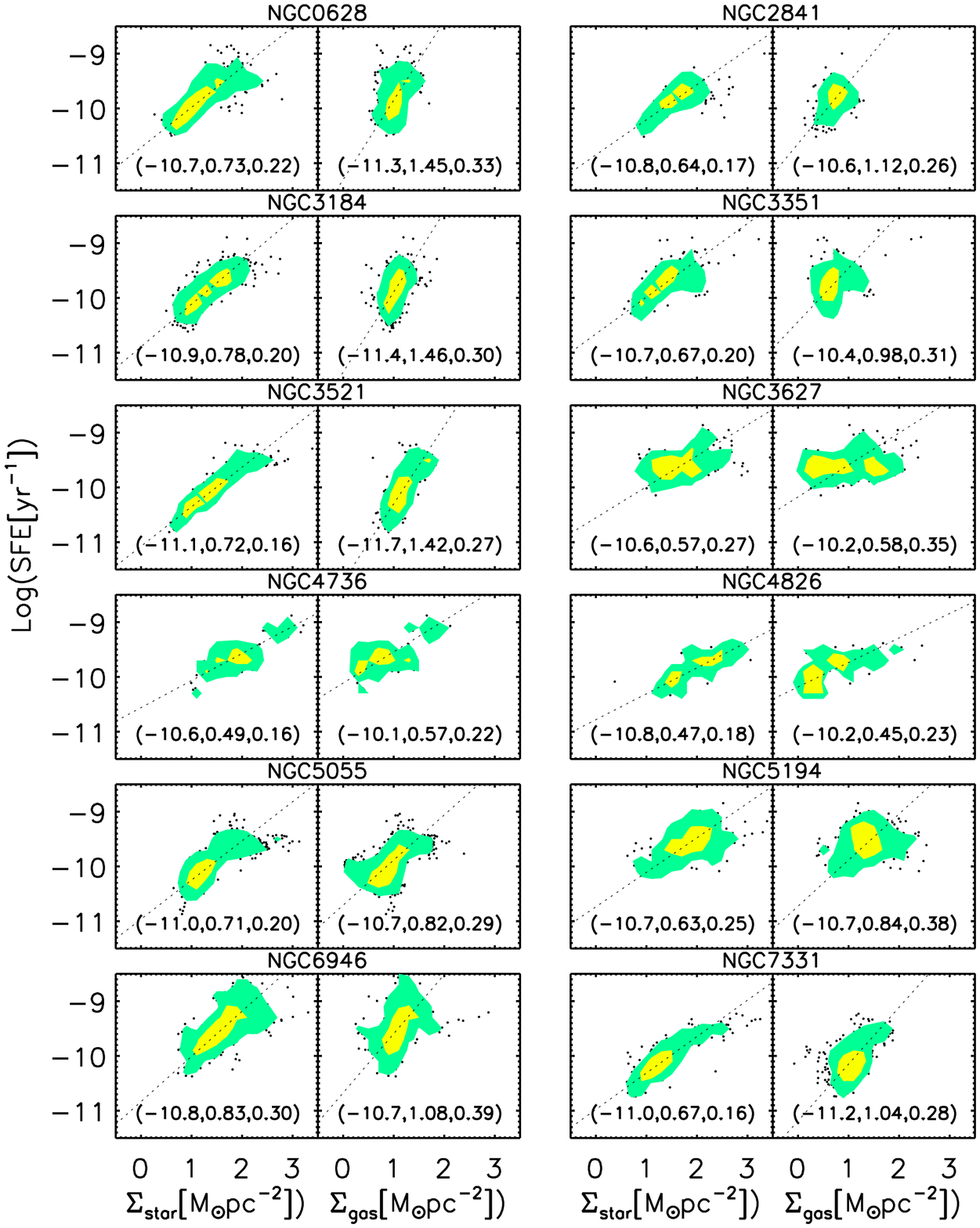}
\caption{\label{region_ind_spiral} The extended Schmidt law at sub-kpc
resolution (the left  side of each panel) compared to  the KS law (the
right  side of  each panel)  in a  sample of  12 spiral  galaxies. The
filled  yellow and  green areas  enclose 50\%  and 90\%  of  the total
datapoints, respectively,  while the dots  are those outside  the 90\%
area.  The numbers listed  in the  parenthesize are  in a  sequence of
interception, slope  and 1-$\sigma$ observed  scatter as given  by the
ordinary least squares bisector  fit \citep{Isobe90}, while the dotted
line is the best fit.}
\end{figure*}

We now compare  the extended Schmidt law to the KS  law in the ability
to  predict the  SFE and  SFR  for various  galaxy types  as shown  in
Figure~\ref{SFE_SBMstar} and Figure~\ref{SFR_MgasMstar}, respectively.
The main difference  of the extended Schmidt law  from the standard KS
law is to  bring the LSB objects back to  the relationship.  While the
KS   law  was  first   defined  for   late-type  galaxies   and  LIRGs
\citep{Kennicutt89,    Kennicutt98a},   LSB    galaxies/regions   show
significant deviations  from it \citep[e.g.][]{Kennicutt98a, Martin01,
Bigiel08, Wyder09, Roychowdhury09}. The median offsets of the LSB from
the    best    fit    are    -0.75    dex    and    -0.71    dex    in
Figure~\ref{SFE_SBMstar}(b)     and     Figure~\ref{SFR_MgasMstar}(b),
respectively,  where  the  best-fit  is  done  by  excluding  LSB  and
early-type objects due to their  apparent offsets.  The cause for this
deviation  is  still  unclear  but  is most  likely  related  to  some
instabilities either gravitationally  or thermally and chemically.  On
the other  hand, these  LSB objects follow  more or less  the extended
Schmidt law with median offsets of -0.16 and -0.08 dex in two figures,
respectively, where the fit is done  for all objects. This is the main
advantage  of the extended  Schmidt law  compared to  the KS  one. The
observed scatter of the extended  Schmidt law is slightly smaller than
that  of  the KS  law  (see  Table~\ref{fit_result}).  For  early-type
galaxies, we  here included objects from studies  of \citet{Wei10} and
\citet{Crocker11}.  They  follow more or  less the KS law  with median
offsets   of    0.30   dex   and   0.32   dex    toward   higher   SFE
(Figure~\ref{SFE_SBMstar}(b)) and SFR (Figure~\ref{SFR_MgasMstar}(b)),
respectively.   The objects  from \citet{Wei10}  show  slightly larger
offsets than \citet{Crocker11}, probably  due to the contribution from
old stellar populations  to the UV emission that  has been included in
their SFR  measurements.  As shown  in Figure~\ref{SFE_SBMstar}(a) and
Figure~\ref{SFR_MgasMstar}(a), these early-type galaxies lie generally
on the extended  Schmidt law with median offsets of  -0.19 dex and -0.26 dex on two
figures, respectively. As  a summary, we  found that the  extended Schmidt law  is a
universal relationship that holds  for various galaxy types especially
including LSB ones that do not follow the KS law.

As  described in  \S~\ref{sample_ind_gal}, different  galaxy apertures
are    used   for    different   galaxy    types.    As    listed   in
Table~\ref{tab_SAMP_COHI}, their  relative sizes to  optical isophotal
radii $R_{25}$ ranges from around 0.2$R_{25}$ for LIRGs and early-type
galaxies  with circumnuclear  star  formation to  2-3$R_{25}$ for  LSB
galaxies  with  widely-distributed  star  formation.   Most  of  these
apertures are indeed defined to enclose the majority of star formation
and gas, which is consistent with the definition of star formation law
that empirically describes  how stars from gas. In  spite of different
apertures,  the extended Schmidt  law depends  little on  the aperture
size  as long  as three  quantities (SFR,  gas and  stellar  mass) are
measured within the same aperture which  is what has been done in this
study.   As shown  in the  below,  galaxies basically  move along  the
relation  without  large  offsets   from  the  best-fit  if  different
apertures  are used.  For  11 LSB  galaxies from  \citet{Wyder09} with
available radial  profiles of SFR,  gas and stellar mass,  we measured
the relative  offset in the Y-axis  from the best-fit  compared to the
aperture ($\sim$2$R_{25}$)  used in Figure~\ref{SFR_MgasMstar}.  Three
apertures  of  0.2$R_{25}$,   0.5$R_{25}$  and  $R_{25}$  are  tested.
Although they move along the relation, the median relative offsets are
only  -0.02  dex, 0.12  dex  and  0.04  dex, respectively,  where  the
positive sign mean  offset toward higher SFRs. For  12 spiral galaxies
with apertures of 1.5$R_{25}$  from \citet{Leroy08}, the median offset
for 0.5$R_{25}$ and $R_{25}$ are -0.1 dex and -0.02 dex, respectively.
For  early-type and ULIRGs,  if we  assume no  star formation  and gas
outside the circumnuclear region, the median offset will be around 0.3
dex and 0.6 dex for the  aperture radius of 0.5 R$_{25}$ and $R_{25}$,
respectively,  which  is  still  with  tolerance  given  the  observed
1-$\sigma$ scatter of  0.5 dex in Figure~\ref{SFR_MgasMstar}. Overall,
we have  found that the extended  Schmidt law changes a  little if the
aperture radius vary several times (3-5).

Accurate stellar mass measurements are  important to the result of the
extended  Schmidt  law.   To  reduce systematic  errors  by  different
studies, we have measured the  masses by our own based on \citet{BC03}
model for almost all objects except for 23 galaxies in \citet{Leroy08}
for which  the median  offset from  our method is  only 0.1  dex.  The
\citet{BC03} model  underestimates the contribution from  AGB stars in
the near-IR and thus  overestimates the near-IR mass-to-light ratio at
ages around  0.1-3Gyr. If the stellar  masses are reduced  by 60\% for
galaxies with  characteristic ages defined  by stellar-mass/SFR around
the above range, the slope  of the extended Schmidt law increases only
by  $\sim$0.01  dex. As  listed  in  Table~\ref{BC_model}, we  adopted
exponentially declining  or constant  star formation history  (SFH). If
the  SFH   is  exponentially  increasing  for   the  high-z  objects
\citep{Maraston10, Papovich10}, the  inferred stellar mass decreases by
about 0.2 dex and the slope of the extended Schmidt law only increases
by 0.01 dex.  We notice  that the photometric coverage of LSB galaxies
is  generally  not   as  good  as  other  types.    While  those  from
\citet{Wyder09} is  essentially based on the  SDSS optical photometry,
the ones from \citet{Leroy08} are based on IRAC 3.6 $\mu$m. The median
offset in the  stellar mass density between the  two subsample is only
0.04  dex, implying  that there  is no  significant bias  in  the mass
estimate based only on either optical or near-IR photometry.

Recently \citet{Schiminovich10} published the HI-based SFE measurement
for a  large sample of  local massive galaxies ($M_{*}$  $>$ 10$^{10}$
M$_{\odot}$).   They  claimed  an  almost  constant  SFE  (10$^{-9.5}$
yr$^{-1}$) in their sample for  a range of log($\Sigma_{\rm star}$) of
2 to 3.3 M$_{\odot}$kpc$^{-2}$.  The detailed comparison to our result
is hampered by the lack of the H$_{2}$ data of their objects, possible
large uncertainties in their  SFR measurements (UV-based ones vs.  our
recombination-line/IR/UV   based  ones)   and  most   importantly  the
aperture-unmatched SFE  measurements relative to those  of the stellar
densities  measured  within the  half  light  radius ($r_{1/2}$).   To
estimate a rough deviation of  their sample from the prediction of our
relation   given    their   median   density,    we   assume   $M_{\rm
HI+H_{2}}/M_{\rm H_{I}}$ $\approx$ 10 within $r_{1/2}$ \citep{Leroy08}
and $M_{\rm  HI}^{\rm tot}/M_{\rm HI}^{r_{1/2}}$  $\approx$ 50 derived
from the THINGS HI  radial profile in \citet{Walter08}.  The resulting
offset is $<$ 0.3 dex.

The  above  studies assume  the  same  CO-to-H$_{2}$ factor  ($\alpha$
value) for all  galaxies, while this factor is  likely to be different
in local ULIRGs and high-z  merging galaxies from normal galaxies.  To
quantify the effect  of different $\alpha$ values in  the relation, we
show the relations in  Figure~\ref{result_diff_a} with $\alpha$ = 3.17
M$_{\odot}$  pc$^{-2}$  (K  km  s$^{-1}$)$^{-1}$ for  normal  galaxies
\citep{Dickman86,  Solomon87, Tacconi08} and  $\alpha$=1.0 M$_{\odot}$
pc$^{-2}$ (K  km s$^{-1}$)$^{-1}$ for merging galaxies  (z=0 LIRGs and
high-z  sub-millimeter   galaxies  in  this   study)  \citep{Downes98,
Tacconi08}.   For  the  KS  law  (Figure~\ref{result_diff_a}(d)),  the
median offsets of  the merging galaxies from the best  fit to all data
points and non-merging galaxies  excluding LSB and early-type ones are
0.48 and 0.81  dex, respectively, which is consistent  with the result
obtained by  \citet{Genzel10} and \citet{Daddi10b}.   For the extended
Schmidt law  (Figure~\ref{result_diff_a}(c)), merging galaxies offsets
of 0.58 and 0.74 dex from the best fit to all objects and star-forming
galaxies, respectively,  which is also  comparable to the case  of the
SFE-$t_{\rm dyn}$  relationships (0.5-0.7 dex  by \citet{Genzel10} and
about  0.3  dex  by  \citet{Daddi10b}).   Thus even  in  the  case  of
different CO-to-H$_{2}$  factors for mergers, the basic  idea that the
SFE is regulated by the stellar density still holds.

\subsection{The Extended Schmidt Law At Sub-Kpc Resolution}\label{res_ind_reg}

\begin{figure*}
\epsscale{1.0}
\plotone{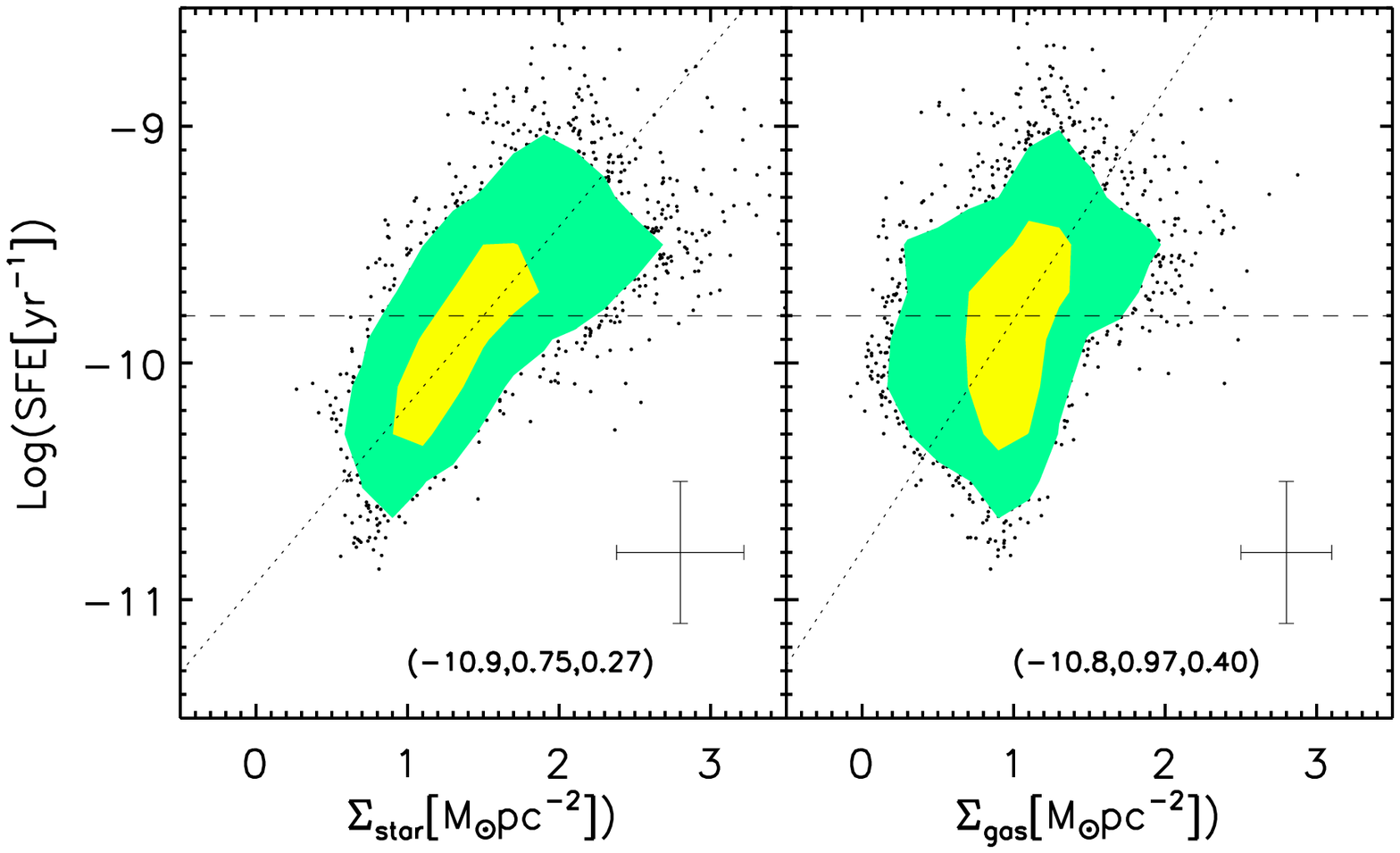}
\caption{\label{region_all_spiral}   The extended Schmidt law at sub-kpc
resolution (the  left side of  each panel) and  the KS law  (the right
side of each panel) of all  sub-kpc regions in 12 spiral galaxies. The
filled  yellow and  green areas  enclose 50\%  and 90\%  of  the total
datapoints, respectively,  while the dots  are those outside  the 90\%
area.  The numbers listed  in the  parenthesize are  in a  sequence of
interception, slope  and 1-$\sigma$ observed  scatter as given  by the
ordinary least squares bisector  fit \citep{Isobe90}, while the dotted
line is the best fit. The horizontal dashed line marks the transition below which
the KS law has a much steeper slope \citep[also see ][]{Bigiel08}. }
\end{figure*}

The result  of the  extended Schmidt law  at sub-kpc resolution  in 12
spiral  galaxies  is  shown  in   the  left  hand  of  each  panel  in
Figure~\ref{region_ind_spiral}. It  clearly indicates that  the SFE is
also a  function of  the stellar mass  density for sub-kpc  regions in
individual spiral galaxies. This demonstrates that the extended Schmidt
law proposed  for global galaxies in  the above section  also works at
sub-kpc resolution. Compared to the KS  law as shown in the right hand
of each  panel, the extended  Schmidt law shows  significantly smaller
scatter.   Within individual  galaxies,  the observed  scatter of  the
extended Schmidt law is on average  1.5 times smaller than that of the
KS law.  Across 12  galaxies, the mean  and standard deviation  of the
slope of  the extended Schmidt  law is 0.66$\pm$0.11 whose  scatter is
three times smaller than that of the KS law (0.98$\pm$0.35). 

Figure~\ref{region_all_spiral} shows the overall trend for all galaxies.  A linear
regression fit (IDL regress.pro) to gas mass, stellar mass and SFR surface densities gives:
\begin{equation}{\label{eq_fit_three}}
 \Sigma_{\rm   SFR} \propto \Sigma_{\rm gas}^{0.80\pm0.01}\Sigma_{\rm star}^{0.63\pm0.01}
\end{equation}  
Again, the exponent of the  stellar mass density is significantly from
zero,  indicating the importance  of this  quantity in  predicting the
SFR.   The  fit to  SFE  vs.  $\Sigma_{\rm  star}$  gives  a slope  of
0.75$\pm$0.01   with  the  ordinary   least  square   bisector  method
\citep{Isobe90}. As shown in Figure~\ref{region_all_spiral}, below SFE
of 10$^{-9.8}$ yr$^{-1}$ (horizontal dashed line), the slope becomes much 
steeper for the KS law, consistent with what found in \citet{Bigiel08}.
On the other hand, such a large deviation is not seen for the extended Schmidt law.
This further suggests the universality of the extended Schmidt law at sub-kpc resolution.

Many previous spatially-resolved studies  of nearby galaxies have also
noticed the trend of SFR as a function of stars. \citet{Ryder94} found
that  the H$\alpha$  emission  spatially follows  the distribution  of
$I$-band stellar emission in  spiral disks.  Their quantitative result
gives      $\Sigma_{\rm      H\alpha}$     $\propto$      $\Sigma_{\rm
I-band}^{0.64\pm0.37}$  within   and  among  galaxies.    In  the  LSB
galaxies, \citet{Hunter98}  also found that the radial  profile of the
SFR follows that  of the stellar mass density but  not the gas density
profile. Several other works have also noticed similar clues about the
relationship    between   existing    stars    and   star    formation
\citep{Brosch98,  Hunter04}.    Recently,  \citet{Leroy08}  have  also
pointed  out a  correlation between  the SFE  and stellar  density but
claimed different slopes (see their Figure 3): a slope of unity in the
HI-dominated  regime  with  1  $<$  log($\frac{\Sigma_{\rm  star}}{\rm
M_{\sun}pc^{-2}}$) $<$  1.9 and a  constant trend (zero slope)  in the
H$_{2}$-dominated   regime    of   log($\frac{\Sigma_{\rm   star}}{\rm
M_{\sun}pc^{-2}}$) $>$  1.9.  We re-analyzed their data  and found two
factors that cause  this inconsistency.  We note that  the unity slope
in that work is not from a fit and the overall variation in the SFE at
1 $<$ log($\frac{\Sigma_{\rm  star}}{\rm M_{\sun}pc^{-2}}$) $<$ 1.9 is
almost a factor of 10, too large to constrain the slope.  A direct fit
to all of their data points gives a slope of 0.65, close to our value.
Also, they do not account for the color gradient in their stellar mass
measurements, which would steepen the intrinsic slope. All these works
together strongly demonstrate  the existence of the SFE  as a function
of the stellar mass at sub-kpc resolution, while ours further indicate
its significantly smaller scatter compared to the KS law.

\section{Discussion}\label{discussion}

\subsection{Test Of Theoretical Models Of The Star Formation Recipe}

\subsubsection{Is the extended Schmidt law just another form of the KS law? }

\begin{figure}
\epsscale{1.0}
\plotone{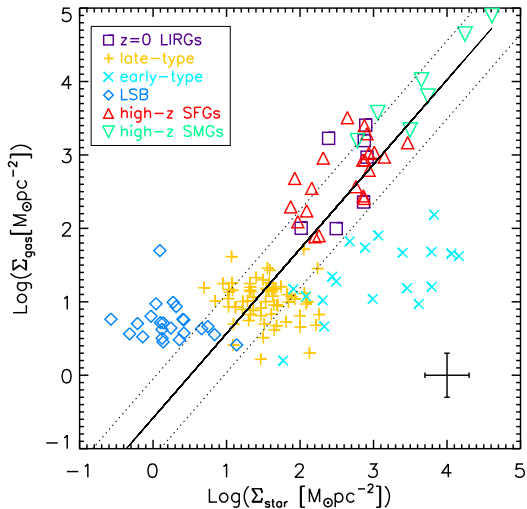}
\caption{\label{Egas_Estar} The stellar mass surface density vs. gas surface density. The 
solid line is the fit to  galaxies excluding LSB and early-type ones. The dotted lines is the
observed 1-$\sigma$ scatter.  }
\end{figure}

Can  the extended  Schmidt  law be  a result  of  the KS  law and  the
relation  between gas  and stellar  densities? Figure~\ref{Egas_Estar}
shows the relation between $\Sigma_{\rm gas}$ and $\Sigma_{\rm star}$.
Excluding the LSB and early-type  objects, the gas density scales with
the  stellar density  with an  observed scatter  of 0.5  dex.   If the
extended Schmidt law  is derived from the KS  law and the $\Sigma_{\rm
gas}$-$\Sigma_{\rm star}$, it  should have a scatter of  0.7 dex given
the scatter of the SFE-$\Sigma_{\rm gas}$ relation of 0.5 dex, whereas
the observed  scatter of the extended  Schmidt law is only  0.4 dex as
listed in  Table~\ref{fit_result}. This suggests  that the KS  law and
$\Sigma_{\rm gas}$-$\Sigma_{\rm  star}$ relations are  not fundamental
relations that drive the SFE-$\Sigma_{\rm gas}$ one. Stronger evidence
comes  from the  LSB galaxies  which do  not follow  either KS  law or
$\Sigma_{\rm  gas}$-$\Sigma_{\rm  star}$ relation  but  do follow  the
SFE-$\Sigma_{\rm gas}$ relation.  In addition, the early-type galaxies
also  seems  offset from  the  $\Sigma_{\rm gas}$-$\Sigma_{\rm  star}$
relation but they  do follow the extended Schmidt  law.  By invoking a
different  physical  parameter  ($\Sigma_{\rm  star}$),  the  extended
Schmidt law presents  another star formation law that  is not a simple
recasting of  the KS relation.  In  the remainder of  this section, we
will test  several simple physical  star formation models in  order to
understand its origin.

\subsubsection{Free-fall In A Star-Dominated Potential}
  
While the KS law can be interpreted as a free-fall in a gas-dominated 
gravitational potential, we note that the extended Schmidt law is consistent with the idea of free-fall
in a stellar potential. The SFR can be expressed as the amount of gas collapsing 
into stars within  a given timescale:
\begin{equation}{\label{ff_eqn}}
 \Sigma_{\rm SFR} = \frac{\eta\Sigma_{\rm gas}}{\tau}
\end{equation}
where $\eta$ gives the fraction of the total gas that collapses into stars and
$\tau$ describes the duration of gas collapse. A free-fall  gas collapse in a 
star-dominated potential has a timescale:
\begin{eqnarray}
\tau_{\rm ff} &=&\frac{1}{4}\sqrt{\frac{3\pi}{2G(\rho_{\rm gas}+\rho_{\rm star})}} \approx \frac{1}{4}\sqrt{\frac{3\pi}{2G\rho_{\rm star}}} 
                =\frac{1}{4}\sqrt{\frac{3\pi{h_{\rm star}}}{G\Sigma_{\rm star}}} \nonumber \\
            &=&3.5\times10^{8} {\rm yr} (\frac{h_{\rm star}}{\rm 1 kpc})^{0.5} (\frac{\rm 1 M_{\odot}pc^{-2}}{\Sigma_{\rm star}})^{0.5}
\end{eqnarray} 
where $\rho_{\rm gas}$, $\rho_{\rm star}$, $\Sigma_{\rm
star}$ and  $h_{\rm star}$  are the gas  mass volume  density, stellar
mass  volume density,  stellar mass  surface density  and  the stellar
scale  height,  respectively.  For  a  self-gravitating stellar  disk,
$\Sigma_{\rm  star}$= 2$\rho_{\rm  star}$$h_{\rm star}$.   This simple
interpretation  predicts   the  observed   power  index  of   0.5  for
$\Sigma_{\rm star}$,  if $\rho_{\rm star}$ $\gg$  $\rho_{\rm gas}$ and
$h_{\rm star}$ = constant.  The comparison of Equation~\ref{ff_eqn} to
the observed relation gives:
\begin{equation}
\eta \approx 2\%
\end{equation}

How  well does  this scenario  represent reality?   How does  the
stellar potential act on the  gas collapse?  As shown above, the above
derivation  assumes  two  conditions  $h_{\rm star}$  =  constant  and
$\rho_{\rm star}$ $\gg$ $\rho_{\rm gas}$. While there is evidence that
the stellar scale height remains  constant with radius within a galaxy
\citep{vanderKruit81,  Fry99},   the  variation  among   galaxies  may
contribute   to   the   scatter   of  the   correlation   under   this
interpretation.  The  condition of $\rho_{\rm  star}$ $\gg$ $\rho_{\rm
gas}$ is true for general high-surface-brightness galaxies but not for
LSB ones as shown in  Figure~\ref{Egas_Estar} which shows that the gas
potential  dominates over  the  stellar  one. If  the  gas density  is
included   in   the  correlation   as   SFE  $\propto$   $(\Sigma_{\rm
gas}+\Sigma_{\rm star})^{0.5}$, the  correlation will not improve, but
instead LSB galaxies will be offset toward the high density end. There is
no room for additional gas self-gravity in the empirical relation.

\subsubsection{Pressure-Regulated H$_{2}$ Formation}

The  prerequisite to  star formation  is formation  of  cold molecular
H$_{2}$. Many works  have highlighted the role of  the stellar gravity
in regulating  H$_{2}$ formation \citep{Elmegreen93,  Wong02, Blitz04,
Blitz06}. We explore here whether the extended Schmidt law actually
reflects the process of H$_{2}$ production from HI. Quantitatively,  the H$_{2}$-to-HI
mass ratio ($R_{\rm mol}$) can be written as a function of pressure with a power index of $\gamma$:
\begin{equation}{\label{H2_HI_P}}
R_{\rm mol}={\rm H}_{2}/{\rm HI} \propto P_{\rm ext}^{\gamma}
\end{equation}
\citet{Blitz04} estimate the external pressure $P_{\rm ext}$ as the mid-plane pressure in an infinite two-fluid
isothermal disk where the gas scale height is much less than the stellar height:
\begin{eqnarray}{\label{Pext_star_gas}}
 P_{\rm ext} &=&  (2G)^{0.5}\Sigma_{\rm gas}v_{\rm gas}[\rho_{\rm star}^{0.5}+(\frac{\pi}{4}\rho_{\rm gas})^{0.5}] \nonumber \\
           &=&  0.84(G\Sigma_{\rm star})^{0.5}\Sigma_{\rm gas}\frac{v_{\rm gas}}{h_{\rm star}^{0.5}} \nonumber \\
           &=&  (272 {\rm cm^{-3} K}) (\frac{\Sigma_{\rm gas}}{\rm M_{\odot} pc^{-2}}) (\frac{\Sigma_{\rm star}}{\rm M_{\odot} pc^{-2}})^{0.5} \nonumber \\
           & &  (\frac{v_{\rm gas}}{\rm km s^{-1}})(\frac{h_{\rm star}}{\rm pc})^{-0.5}(k)
\end{eqnarray}
where $\Sigma_{\rm gas}$ is  the mid-plane gas surface density, $v_{\rm
gas}$ is the  vertical velocity dispersion of the  gas disk, $\rho_{\rm
star}$ is the mid-plane stellar volume density, $\rho_{\rm gas}$ is the
mid-plane  gas volume  density,  $\Sigma_{\rm star}$ is  the mid-plane  stellar
surface density, $h_{\rm star}$ is the stellar scale height and $k$ is
the Boltzmann constant. By assuming constant $v_{\rm gas}$ for the gas disk and constant $h_{\rm
star}$ for the stellar disk, the Equation~\ref{H2_HI_P} has been
demonstrated    observationally    with    ${\gamma}$    around    1.0
\citep{Blitz06}.  Under  this  assumption  and that  stars  form  from
molecular gas, the star formation prescription can be written as:
\begin{eqnarray}
\Sigma_{\rm SFR} &=& \frac{\eta_{\rm H_{2}}\Sigma_{\rm H_{2}}}{\tau}=\frac{\eta_{\rm H_{2}}}{\tau}\frac{R_{\rm mol}}{1+R_{\rm mol}}\Sigma_{\rm gas} \nonumber \\
               &=& \frac{\eta_{\rm H_{2}}}{\tau}\frac{(P_{\rm ext}/P_{0})^{\gamma}}{1+(P_{\rm ext}/P_{0})^{\gamma}}\Sigma_{\rm gas}
\end{eqnarray}
where $\eta_{\rm H_{2}}$ is the  fraction of the molecular gas that is
locked in stars, $\tau$ is the timescale for the collapse of molecular
clouds to stars, $\gamma$$=$0.92 and $P_{0}/k$ is 4.3$\times$10$^{4}$
cm$^{-3}$  K  as  given  by  observation in  \citet{Blitz06}. The $\frac{\eta_{\rm H_{2}}}{\tau}$ is
observed   to   be  a constant \citep[e.g.][]{Gao04, Wu05, Leroy08, Genzel10}.
 We now  discuss $\Sigma_{\rm SFR}$ in two extreme
pressure regimes:

HI-dominated galaxies ($P_{\rm ext}/P_{0}$ $\ll$ 1, $R_{\rm mol}$ $\ll$ 1): The above equation
gives
\begin{equation}
\Sigma_{\rm SFR}  \propto (P_{\rm ext}/P_{0})^{\gamma}\Sigma_{\rm gas}\propto\Sigma_{\rm star}^{0.5\gamma}\Sigma_{\rm gas}^{1.0+\gamma}
\end{equation}
for constant $h_{\rm star}$ and  $v_{\rm gas}$. At $\gamma=$0.92, the above equation predicts roughly the same
power index for the $\Sigma_{\rm star}$ as we observe but almost two times larger for that of $\Sigma_{\rm gas}$.

H$_{2}$-dominated galaxies ($P_{\rm ext}/P_{0}$ $\gg$ 1, $R_{\rm mol}$ $\gg$ 1): It is obvious in this
regime there is no dependence of $\Sigma_{\rm SFR}$ on $\Sigma_{\rm star}$, inconsistent with the 
extended Schmidt law for $H_{2}$-dominated circumnuclear star-forming regions and LIRGs.

\subsubsection{Pressure-Supported Star Formation}

The scenario of pressure-supported star formation assumes that star formation
is regulated by the pressure balance between gas collapse and feedback from stars
 \citep{Thompson05},  i.e.,  weak stellar
feedback leaves the  gas  collapse unimpeded,  resulting  in enhanced  star
formation, which in turn increases  the feedback to prevent the further
gas collapse;  strong feedback prevents the efficient gas
collapse, which  lowers the  amount of newly-formed  stars and thus decreases
the feedback strength. Quantitatively, we have
\begin{equation}
P_{\rm ext} = P_{\rm SFR}
\end{equation}

\citet{Thompson05} estimates the total pressure from star formation as a sum
of the supernovae feedback and radiation pressure:
\begin{eqnarray}
P_{\rm SFR} & \approx & (P_{\rm SN}+P_{\rm RP}) \approx (5n_{1}^{-1/4}E_{51}^{13/14}+1)P_{\rm RP} \nonumber \\
          & \approx & (5n_{1}^{-1/4}E_{51}^{13/14}+1){\epsilon}c\Sigma_{\rm SFR}
\end{eqnarray}
where $P_{\rm SN}$ is the pressure from supernova, $P_{\rm RP}$ is the 
radiation pressure from massive stars, $n_{1}$ is the density of the 
interstellar medium  in the  unit of 1 cm$^{-3}$, $E_{51}$ is the supernova
energy in the unit of 10$^{51}$ ergs, $\epsilon$ is the conversion efficiency
from the stellar mass into radiation ($\epsilon$ $\sim$ 10$^{-3}$ for a Salpeter
initial mass function) and $c$ is the speed of light. In general quiescent galaxies, the
ISM density is low and the pressure is dominated by supernova while in 
LIRGs the radiation pressure starts to become important or even dominates.

Using the Equation~\ref{Pext_star_gas}
for $P_{\rm ext}$, we have
\begin{eqnarray}
\frac{\Sigma_{\rm SFR}}{\rm M_{\odot}yr^{-1}pc^{-2}} &=& \frac{1.9\times10^{-10}}{(5n_{1}^{-1/4}E_{51}^{13/14}+1)}(\frac{\Sigma_{\rm gas}}{\rm M_{\odot} pc^{-2}}) \nonumber \\
      & & (\frac{\Sigma_{\rm star}}{\rm M_{\odot} pc^{-2}})^{0.5}(\frac{v_{\rm gas}}{\rm km s^{-1}})(\frac{h_{\rm star}}{\rm pc})^{-0.5}
\end{eqnarray}
Comparing to the observed correlation (Equation~\ref{SFE_star_model}), the
above equation produces not only the exact power indices for both $\Sigma_{\rm star}$ and $\Sigma_{\rm gas}$ but also
a similar constant. The caveat to this explanation is again offered by the LSB galaxies 
where the gas gravity cannot be neglected compared to the stellar term as shown in 
Figure~\ref{Egas_Estar} and as discussed above.


\subsubsection{Summary: Causal or Casual}

We compare  the extended Schmidt  law to some  physical star-formation
models   including   gas   free-fall   in   the   stellar   potential,
pressure-regulated  H$_{2}$   formation  and  pressure-supported  star
formation.  All  of them  invoke roles of  the existing stars  in star
formation through stellar gravity on gas, and the first and third ones
predict   not  only   the  same   power  indices   but   also  similar
normalizations to those observed ones.  However, as pointed above, the
assumption that stars dominate the mass seems unreasonable for the
LSB galaxies  at least in  case of hydrostatic  equilibrium.  However,
the stellar gravity can affect gas motion critically in configurations
where stars  show spatial and  velocity differences from gas,  such as
stellar  bars.   An example  of  this may  be  seen  in the  numerical
simulation    of   gas-dominated   merging    galaxies   \citep{Hopkins09a,
Hopkins09b}.   \citet{Springel05} and  \citet{Robertson06}  have shown
that gas-rich  mergers result in  disk galaxies instead  of elliptical
galaxies.  The reason for this  is not just that gas-rich mergers have
too much gas to consume but also that they lack existing stars.  Stars
are collisionless  and can relax its orbits  violently during merging.
Gas, on the  other hand, is collisional and  cannot relax
rapidly, requiring angular momentum  to be removed in order to form
stars \citep{Hopkins09a}.  With non-axisymmetric distortion as induced
by the secondary galaxy, the  gravity of these stars thus provides the
most  efficient  way to  remove  the  gaseous  angular momentum.   Its
efficiency  far exceeds  those of  shock compression,  gravity  of the
secondary   galaxy    and   self   gravity   of    the   gas   itself.
\citet{Hopkins09b} have  derived an analytic  expression that captures
the role of existing  stars: $f_{\rm starburst}/f_{\rm gas}$ $\propto$
$f_{\rm star}$, where $f_{\rm starburst}$ is the fraction of the total
mass that forms stars, $f_{\rm  gas}$ is initial gas mass fraction and
$f_{\rm star}$  is initial stellar mass  fraction. Determining whether
similarly critical influence of stars  on gas flow and processing also
obtains  in LSB and  normal galaxies  will probably  require extensive
simulations.

Since the physical models of  star formation do not translate directly
into the extended Schmidt law, one is not free to interpret the latter
as  a  causal  formula   suggesting  that  stellar  gravity  regulates
SFE. Instead,  $\Sigma_{\rm star}$ may  be a proxy for  other physical
parameters  or a  combination thereof,  signifying regulation  by more
subtle  or complex  physics.   $\Sigma_{\rm star}$  may represent  the
total kinetic and/or radiation energy  dumped into the ISM by stars or
the total metal enrichment over the galaxy's history. For example, the
metal  abundance is  critical in  ISM  cooling and  formation of  dust
grains on  which H$_{2}$ can  form efficiently. Theoretical  models do
confirm the significant deviation of the KS law in the low metallicity
environment \citep{Krumholz09, Gnedin10, Papadopoulos10}.

\subsection{Implications For the Star-Forming Main Sequence}

\begin{figure}
\epsscale{1.2}
\plotone{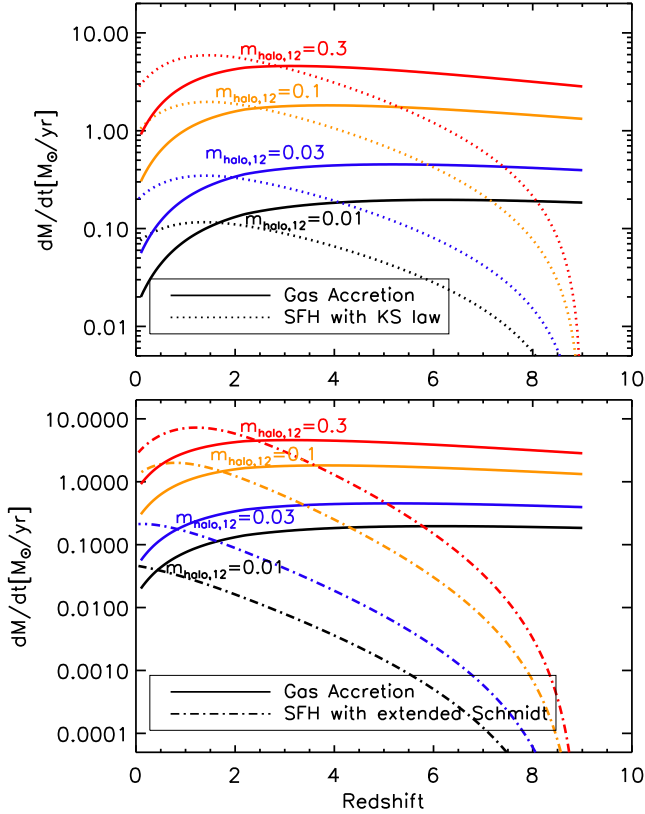}
\caption{\label{evl_model} Upper: The  SFR histories (dotted lines) as
predicted by the KS law for given gas accretion histories (solid lines
with  the same  colors) in  four different  halo masses  where $m_{\rm
halo,  12}$  is the  halo  mass  at $z$=0  in  the  unit of  10$^{12}$
M$_{\odot}$.  Lower: The SFR histories with the extended Schmidt law.}
\end{figure}

\begin{figure}
\epsscale{1.2}
\plotone{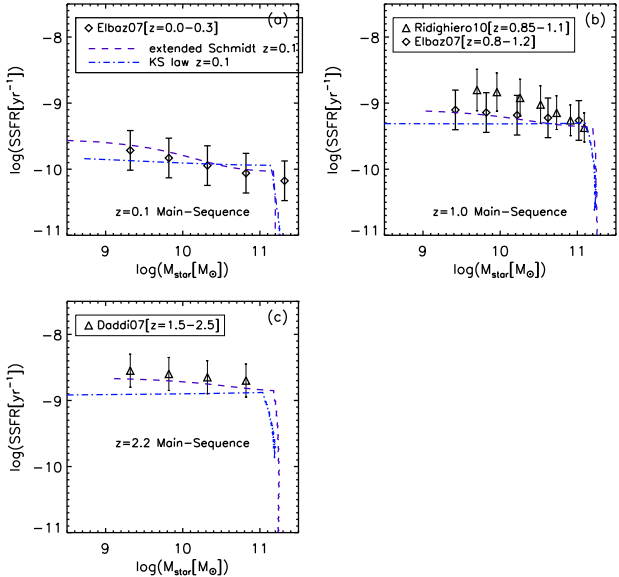}
\caption{\label{main_sequence} The observed star-forming main-sequence
 from z=0 to z=2  compared to the predictions by implementing
the extended Schmidt law (dashed lines) and the  KS law (dotted-dashed
lines) into the  analytic model of gas accretion  in the $\Lambda$ CDM
cosmology. }
\end{figure}

\begin{figure}
\epsscale{1.2}
\plotone{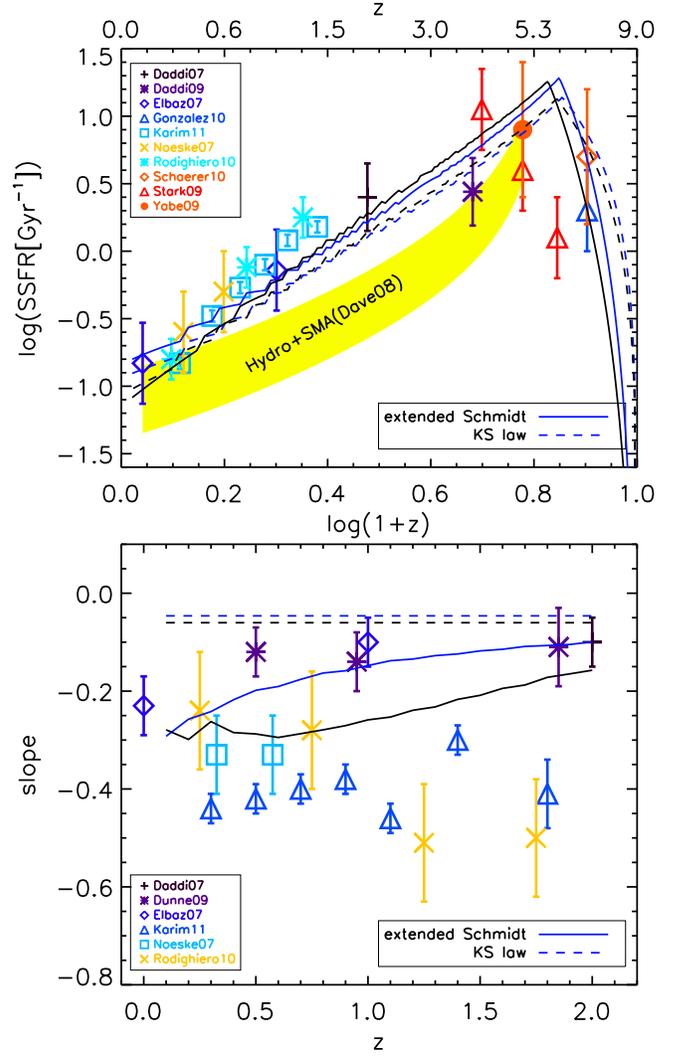}
\caption{\label{ssfr_z}  Upper panel:  The observed  evolution  of the
SSFR  (specific  star  formation  rate) for  $M_{\rm  star}$=10$^{10}$
M$_{\odot}$ compared to the predictions by the extended Schmidt (solid
lines) and  KS laws (dashed lines).  The data are  taken directly from
the literature except for \citet{Stark09} where we further correct the
extinction  according   to  the  luminosity-   and  redshift-dependent
extinction correction  curve of \citet{Bouwens09}.  ``SMA'' stands for
the semi-analytic  model. Lower Panel:  The evolution of  the observed
slope  (specific  SFR  $\propto$  $M_{\rm  star}^{\rm  slope}$)  as  a
function of the redshift from  different studies. The solid and dashed
line show the result for the extended and KS laws, respectively, while
the  blue and  black  color  indicate the  galaxy  size evolution  of
$\propto$   $(H(z=4)/H(z))$   and   $\propto$   $(H(z=4)/H(z))^{0.4}$,
respectively.}

\end{figure}

We here discuss the implication of the extended Schmidt law for galaxy
formation and  evolution with focus on the main sequence of star-forming galaxies.
Studies have  shown that  stars form mainly  in the  blue star-forming
galaxies while there is little star formation in red galaxies.  Such a
bi-modality  has been  well established  through  various observations
\citep[e.g.][]{Blanton03}.  While red galaxies show a relation between
the UV/optical  color and  luminosity, the SFRs  of blue  galaxies are
found  to correlate  with  stellar masses,  a relationship  with a
slope a bit  below unity and a small dispersion  of $\lesssim$ 0.3 dex
\citep{Brinchmann04,  Elbaz07,  Daddi07,  Zheng07, Noeske07a,  Chen09,
Oliver10, Rodighiero10}.  This so-called main sequence of star-forming
galaxies and its evolution  have provided important constraints on the
mechanism driving the rapid evolution  of the cosmic SFR density.  For
example, an evolving  stellar IMF is able to  explain the evolution of
the  main sequence  as  proposed by  \citet{Dave08}.   A more  general
interpretation would  attempt to  quantify the star  formation history
(SFH), since
\begin{eqnarray}\label{eq_MS}
 M_{\rm star}(z_{0}) & \propto & \int_{z_{\rm form}}^{z_{0}}SFR(z)dz=SFR(z_{0})\int_{z_{\rm form}}^{z_{0}}\frac{SFR(z)}{SFR(z_{0})}dz \nonumber \\
                  & = & SFR(z_{0})\int_{z_{\rm form}}^{z_{0}}SFH(z)dz
\end{eqnarray}  where $z_{0}$ is  the observed  redshift of  a galaxy,
$z_{\rm form}$ is the redshift where the galaxy starts to form and SFH
is the star-formation history  normalized by the current SFR.  Current
numerical  simulations  and   analytic  models  have  difficulties  in
producing  the observed  slope  below unity  without  invoking ad  hoc
mechanisms   to   delay   star   formation   in   low   mass   systems
\citep{Noeske07b, Dave08,  Bouche10}.  This can be  seen from Equation
\ref{eq_MS}.  If all  galaxies have a similar smooth  shape of SFH and
form at the same redshift,  then Equation \ref{eq_MS} gives a slope of
unity. To have  a shallower slope, the integral  of the SFH normalized
by the  current SFR needs to be  smaller for a lower  mass galaxy (see
Equation \ref{eq_MS}), for example, $z_{\rm  form}$ can be lower for a
lower   mass  galaxy.   This   can  be   also  rephrased   as  shorter
characteristic  star-formation   timescale  or  late   onset  of  star
formation in  a lower mass  galaxy. However, in  numerical simulations
and  semi-analytic models,  the  gas accretion  is  determined by  the
well-known dark  matter halo growth in $\Lambda$  CDM cosmology.  Thus
SFH is not  a free parameter to adjust  \citep{Dave08}.  The growth of
the dark  matter halo  follows $\dot{M}_{\rm halo}$  $\propto$ $M_{\rm
halo}^{s}$  with  $s$  above unity  \citep[e.g.][]{Neistein08},  which
would result  in a slope of  the main-sequence above unity  if the SFR
follows  that of  the dark  matter  halo growth.   In current  models,
ad-hoc mechanisms  are thus  proposed to delay  star formation  in low
mass galaxies, such  as, a very strong feedback or  a mass floor below
which  the gas  cannot be  accreted  \citep[e.g.][]{Dave08, Bouche10}.
Compared  to  the KS  law  that is  now  widely  invoked in  numerical
simulations  and  semi-analytic   models,  the  extended  Schmidt  law
indicates  a slow  SFR increase  at  early times  due to  the lack  of
existing  stars and  fast  evolution at  late  times for  a given  gas
accretion history.  This introduces naturally a {\it delayed} onset of
star formation  in a low-mass galaxy,  which is the  key to explaining
the  star-forming main  sequence with  a  slope below  unity.  In  the
following, we show quantitatively that the delayed star formation in a
low mass system is a natural result of star formation that is governed
by the extended Schmidt law.

We  follow exactly \citet{Bouche10} to  construct the growth  of the
dark  matter  halo and  gas  accretion.   For  a given  gas  accretion
history, star  formation occurs following  either the extended Schmidt
 law or the KS law. To apply these two laws, we assume the evolution
of the half-light radius ($R_{1/2}$) from \citet{Papovich10}.
The numerical calculations include:

(1) The dark matter halo growth rate follows:
\begin{equation}\label{det_mhalo_acc} 
\dot{M}_{\rm halo} = 510M_{\rm halo,12}^{s} ((1+z)/3.2)^{t} \mbox{ M$_{\odot}$/yr}
\end{equation}
where $M_{\rm halo,12}=M_{\rm halo}/10^{12}M_{\odot}$, $s$=1.1 and $t$=2.2.

(2) The gas accretion rate is given by:
\begin{equation}\label{det_mgas_acc} 
\dot{M}_{\rm gas, in} = \epsilon_{\rm in}f_{\rm b}\dot{M}_{\rm halo}
\end{equation}
Similar to \citet{Bouche10}, $f_{\rm b}$ is the baryonic fraction of 0.18. 
$\epsilon_{\rm in}$ is the accretion efficiency that is equal to 0.7 at 
z $>$ 2.2 where cold accretion mode dominates. 
Due to accumulation of stars and hot gas, the accretion efficiency must decrease with time.
For $z$ $<$ 2.2, \citet{Bouche10} simply assumed $\epsilon_{\rm in}$=f(z)$\times$0.7 where $f(z)$ is a linear
function of time with f(2.2)=1 and f(0)=0.5. For $M_{\rm halo}$ above 10$^{12}$ M$_{\odot}$ where the cold mode
accretion is not important, $\epsilon_{\rm in}$=0 as discussed by \citet{Bouche10}. However, we do not introduce a 
low-mass floor ($M_{\rm halo}$=10$^{11}$ M$_{\odot}$) below which $\epsilon_{\rm in}$=0, which is used to
fit the observed main-sequence as shown  by \citet{Bouche10}. This is the main difference
of our model besides adopting the extended Schmidt law.

(3) At each redshift, the net gas accretion is given by
\begin{equation}\label{det_mgas_tt}
 \dot{M}_{\rm gas} = \dot{M}_{\rm gas, in} - (1-R){\times}SFR-\dot{M}_{\rm gas, out}
\end{equation}
where  $R$ is  the  recycled gas  fraction  and equal  to  0.52 for  a
Chabrier IMF in this  study.  $\dot{M}_{\rm gas, out}$=a$\times$SFR is
the gas  outflow where $a$ is set  to be zero by  assuming the outflow
eventually falls  back to form  stars. Unlike the above  two equations
(\ref{det_mhalo_acc}, \ref{det_mgas_acc}) that  are solely determined by
the dark  matter growth, this equation will  produce different results
for different star formation laws.

(4) Star formation follows either the extended Schmidt law:
\begin{eqnarray}\label{SFE_star_model}
\frac{\mbox{SFE}}{\mbox{yr$^{-1}$}} = 10^{-10.28}(\frac{\Sigma_{\rm star}}{\rm M_{\odot}pc^{-2}})^{0.48}
\end{eqnarray}
or the KS law (using the exponent from the fit in this study):
\begin{equation}\label{ks_model}
  \frac{\Sigma_{\rm SFR}}{\rm M_{\odot}/yr/pc^{2}}=10^{-9.90}(\frac{\Sigma_{\rm gas}}{\rm M_{\odot}/pc^{2}})^{1.38}
\end{equation}
where  $\Sigma=\frac{0.5M}{{\pi}R_{1/2}^{2}}$. Following \citet{Papovich10}, the half light radius $R_{1/2}(z)$ is given by
\begin{equation}\label{size_evl}
 \frac{R_{1/2}}{\rm kpc} = 1.7 \frac{H(z=4)}{H(z)}
\end{equation}
where $H(z=4)$=430 km/s/Mpc.

The  above six equations  are solved  numerically with  the formation
$z_{\rm form}$=9  where the  initial gas and  stellar mass are  set to
zero, and a series of initial dark matter masses are assumed. Examples
of gas accretion and SFR histories are shown in Figure~\ref{evl_model}
for  both  KS  and extended  Schmidt  laws.   With  the KS  law,  star
formation  responds  only to  the  accumulated  gas  and thus  quickly
reaches the state where the SFR follows more or less the gas accretion
history  at  later  times  \citep[e.g.][]{Papovich10}.   As  shown  in
Figure~\ref{evl_model}, the  SFHs of different mass  systems are quite
similar  at $z$$<$7  with similar  peak  redshift and  slopes at  both
sides. This implies a  unity slope of star-forming main sequence as
indicated by  Equation~\ref{eq_MS}.  On the other hand,  the SFH given
by the extended  Schmidt law slowly increases at early  time and it is
slower for lower  mass systems simply as a  result of smaller existing
stellar populations.  The  SFHs peak at lower redshift  for lower mass
systems as shown in Figure~\ref{evl_model}, which naturally introduces
a delay mechanism that is  required to explain the main sequence.  The
quantitative  comparisons to the  observed main  sequences in  term of
SSFR (specific star  formation rate) vs.  M$_{\rm star}$  are shown in
Figure~\ref{main_sequence}. While both  relations produce more or less
the correct normalizations, the KS law never produces a negative slope
while the  prediction by the  extended Schmidt law is  consistent with
the  observed  data from  $z$=0  all  the way  up  to  $z$=2.  Such  a
consistency is reached without  introducing ad hoc mechanisms to delay
star formation in low mass  systems, unlike other studies based on the
KS  law   \citep{Noeske07b,  Dave08,  Bouche10}.   Figure~\ref{ssfr_z}
collects  current  studies  of   the  SSFR  evolution  including  both
normalization  at log$M_{\rm  star}/M_{\odot}$=10 and  the  slope.  In
general, these two  relations (extended Schmidt and KS  laws), as well
as the numerical  simulations, all predict the rapid  evolution of the
SSFR,  which reflects  the  gas accretion  history.   Among them,  the
extended Schmidt  law produces  the best match  to the  observed data,
although the discrepancy still exists  at high $z$ ($z$$>$3). As shown
in the  lower panel of Figure~\ref{ssfr_z},  although different slopes
of the main  sequence have been observed by  different studies, all of
them are negative.  It is  clearly shown that the extended Schmidt law
produces much  more consistent values with the  observed data compared
to the KS  law.  The above result about the predicted  slope by the KS
law depends little on the  numerical values of the physical parameters
invoked in Equation \ref{det_mhalo_acc}-\ref{size_evl}, except for the
$s$ parameter that  is unlikely to be below unity.  A steeper slope by
the extended  Schmidt law can be  produced if the  galaxy size evolves
slower ($\propto$  $(H(z=4)/H(z))^{0.4}$) shown as black  lines in the
figure.

Recently, \citet{Dutton10} have re-produced the observed main sequence
through the  semi-analytical model with  the star formation  recipe of
the pressure-regulated H$_{2}$  formation. The resulting two-power-law
star formation relation  has the same slope as the KS  one at the high
density  regime but a  much steeper  slope (2.84)  at the  low density
end. This further indicates that the pure KS law over-predicts the SFR
at  the early  stage of  galaxy  evolution.  Without  imposing ad  hoc
mechanisms to delay star formation  in a low mass system, the extended
Schmidt law does provide a new way to understand the star-forming main
sequence and its evolution.

\section{Conclusions}\label{paper_conclusion}

(1) We  demonstrate empirically the  existence of a  tight correlation
between    the    star    formation    efficiency    (SFE=$\Sigma_{\rm
SFR}$/$\Sigma_{\rm gas}$)  and the stellar  mass density ($\Sigma_{\rm
star}$),  referred as the  extended Schmidt  law. The  correlation was
derived  by  looking  for  the  dependence of  $\Sigma_{\rm  SFR}$  on
$\Sigma_{\rm gas}$ and  $\Sigma_{\rm star}$.  It has a  power index of
0.48$\pm$0.04 and holds over 5 orders of magnitude in the stellar mass
density    for    various   types    of    galaxies   including    the
low-surface-brightness (LSB) ones  that deviate significantly from the
Kennicutt-Schmidt law.

(2)  We further show  that the  extended Schmidt  law also  applies to
spatially resolved  regions at sub-kpc  resolution. In a sample  of 12
spiral  galaxies, the  extended Schmidt  law  not only  holds for  LSB
regions but also shows significantly smaller scatters, a factor of 1.5
and 3  smaller within and  across galaxies, respectively,  compared to
the Kennicutt-Schmidt law.

(3) The extended Schmidt  law may suggest a critical role for
existing  stellar  populations in  on-going  star formation  activity.
Alternatively, it may be a manifestation of more complex physics where
$\Sigma_{\rm star}$ is a proxy  for other variables or processes.  The
comparison  of the  extended Schmidt  law to  several  simple physical
models indicates that models of gas free-fall in stellar gravitational
potential and  pressure-supported star formation produce  not only the
same  power index  but also  a similar  normalization.   However, this
success   is  limited   to  some   cases,  and   the   exact  physical
interpretation of the extended Schmidt law needs further exploration.

(4) By applying this extended Schmidt  law to an analytic model of gas
accretion in $\Lambda$ CDM cosmology, the observed main sequence of star-forming
galaxies is well  reproduced in the model without the  need for ad hoc
mechanisms to delay star formation in low mass systems.

\section{Acknowledgment}

We  thank the  anonymous  referee for  the  detailed and  constructive
comments.  We also  thank Daniel Dale, Leslie K.  Hunt, Eva Schinnerer
and Bruce G  Elmegreen for careful reading and  comments.  The work is
supported through the Spitzer 5MUSES Legacy Program 40539. The authors
acknowledge support by NASA through awards issued by JPL/Caltech. This
work was based on observations  made with the Spitzer Space Telescope,
which is  operated by  JPL/Caltech under a  contract with  NASA.  This
research has  made use of  the NASA/IPAC Extragalactic  Database (NED)
which  is  operated  by  the  Jet  Propulsion  Laboratory,  California
Institute of Technology, under  contract with the National Aeronautics
and Space Administration.   Funding for the SDSS and  SDSS-II has been
provided  by  the  Alfred   P.  Sloan  Foundation,  the  Participating
Institutions, the National Science Foundation, the U.S.  Department of
Energy,  the  National   Aeronautics  and  Space  Administration,  the
Japanese  Monbukagakusho,  the  Max  Planck Society,  and  the  Higher
Education  Funding   Council  for  England.  The  SDSS   Web  Site  is
http://www.sdss.org/.

\clearpage

\LongTables 
\begin{landscape}
\begin{deluxetable}{lllllllllllllllllllllllll}
\tabletypesize{\scriptsize}
\tablecolumns{15}
\tablecaption{\label{tab_SAMP_COHI} Samples with CO and HI data}
\tablewidth{0pt}
\tablehead{ \colhead{name} & \colhead{type} & \colhead{red}  & \colhead{Dist.} & \colhead{Aperture} & \colhead{Area}  & \colhead{log($\Sigma_{\rm SFR}$)} &
            \colhead{log($\Sigma_{\rm gas}$)} & \colhead{Ref$_{\rm sfr, gas}$}  &  \colhead{log($\Sigma_{\rm star}$)}  & Band$_{\rm stellar-mass}$  & Ref$_{\rm star}$ \\
            \colhead{}     & \colhead{}     & \colhead{}  & \colhead{[Mpc]}  & \colhead{}      & \colhead{[kpc$^{2}$]} & \colhead{[M$_{\odot}$/yr/kpc$^{2}$]}  & 
            \colhead{[M$_{\odot}$/pc$^{2}$]}   &  \colhead{}     &  \colhead{[M$_{\odot}$/pc$^{2}$]} \\
            \colhead{(1)} & \colhead{(2)} & \colhead{(3)}           &  \colhead{(4)}            & \colhead{(5)}     
          & \colhead{(6)}  & \colhead{(7)} & \colhead{(8)}  & \colhead{(9)} & \colhead{(10)} & \colhead{(11)} & \colhead{(12)} \\
}
\startdata
NGC0224              & late-type  &  -0.001001 & 0.78       & R$_{25}$                                      & 1106.29    & -3.31      & 0.81       & 1   & 2.00       & IRAC3.6                        & 18 \\
NGC0598              & late-type  &  -0.000597 & 0.80       & R$_{25}$                                      & 133.25     & -2.65      & 1.16       & 1   & 1.58       & IRAC3.6                        & 18 \\
NGC0628              & late-type  &   0.002192 & 7.30       & 1.5$R_{25}$                                   & 764.54     & -2.98      & 0.90       & 2   & 1.22       & IRAC3.6                        & 2 \\
NGC0772              & late-type  &   0.008246 & 35.00      & R$_{25}$                                      & 4231.16    & -3.02      & 1.00       & 1   & 1.53       & nuv,U,B,V,I,J,H,K              & 18 \\
NGC1058              & late-type  &   0.001728 & 9.37       & R$_{25}$                                      & 52.65      & -2.38      & 0.83       & 1   & 1.33       & B,V,I,J,H,K                    & 18 \\
NGC1569              & late-type  &  -0.000347 & 2.20       & R$_{25}$                                      & 2.02       & -0.98      & 1.45       & 1   & 2.24       & nuv,U,B,V,J,H,K                & 18 \\
NGC2336              & late-type  &   0.007352 & 3.50       & R$_{25}$                                      & 38.86      & -2.10      & 1.00       & 1   & 1.22       & U,B,V,J,H,K                    & 18 \\
NGC2403              & late-type  &   0.000437 & 3.20       & 1.5$R_{25}$                                   & 376.68     & -2.99      & 0.93       & 2   & 1.12       & IRAC3.6                        & 2 \\
NGC2841              & late-type  &   0.002128 & 14.10      & 1.5$R_{25}$                                   & 1425.31    & -3.28      & 0.96       & 2   & 1.65       & IRAC3.6                        & 2 \\
NGC2976              & late-type  &   0.000010 & 3.60       & 1.5$R_{25}$                                   & 102.07     & -3.07      & 0.41       & 2   & 1.09       & IRAC3.6                        & 2 \\
NGC3031              & late-type  &  -0.000113 & 3.50       & R$_{25}$                                      & 402.25     & -2.68      & 0.97       & 1   & 2.08       & fuv,nuv,B,V,R,I,J,H,K,IRAC1    & 18 \\
NGC3077              & late-type  &   0.000047 & 3.80       & 1.5$R_{25}$                                   & 63.62      & -2.87      & 1.30       & 2   & 1.50       & IRAC3.6                        & 2 \\
NGC3184              & late-type  &   0.001975 & 11.10      & 1.5$R_{25}$                                   & 1000.98    & -3.05      & 0.75       & 2   & 1.30       & IRAC3.6                        & 2 \\
NGC3198              & late-type  &   0.002212 & 13.80      & 1.5$R_{25}$                                   & 1194.59    & -3.11      & 1.04       & 2   & 1.02       & IRAC3.6                        & 2 \\
NGC3310              & late-type  &   0.003312 & 17.50      & R$_{25}$                                      & 249.96     & -1.32      & 1.26       & 1   & 1.07       & U,B,V,J,H,K                    & 18 \\
NGC3338              & late-type  &   0.004343 & 25.11      & R$_{25}$                                      & 1462.36    & -2.74      & 0.93       & 1   & 1.12       & U,B,V,J,H,K                    & 18 \\
NGC3351              & late-type  &   0.002595 & 10.10      & 1.5$R_{25}$                                   & 794.23     & -2.93      & 0.51       & 2   & 1.50       & IRAC3.6                        & 2 \\
NGC3368              & late-type  &   0.002992 & 9.72       & R$_{25}$                                      & 265.96     & -2.73      & 1.00       & 1   & 2.10       & nuv,U,B,V,J,H,K                & 18 \\
NGC3486              & late-type  &   0.002272 & 10.55      & R$_{25}$                                      & 373.83     & -2.64      & 1.01       & 1   & 0.85       & nuv,U,B,V,J,H,K                & 18 \\
NGC3521              & late-type  &   0.002672 & 10.70      & 1.5$R_{25}$                                   & 1176.28    & -2.75      & 1.08       & 2   & 1.63       & IRAC3.6                        & 2 \\
NGC3627              & late-type  &   0.002425 & 9.30       & 1.5$R_{25}$                                   & 1365.72    & -2.79      & 0.22       & 2   & 1.46       & IRAC3.6                        & 2 \\
NGC3631              & late-type  &   0.003856 & 24.30      & R$_{25}$                                      & 832.50     & -1.91      & 1.20       & 1   & 1.32       & B,V,J,H,K                      & 18 \\
NGC3675              & late-type  &   0.002568 & 18.50      & R$_{25}$                                      & 793.79     & -2.19      & 1.04       & 1   & 1.93       & B,I,J,H,K                      & 18 \\
NGC3726              & late-type  &   0.002887 & 13.50      & R$_{25}$                                      & 354.09     & -2.46      & 1.16       & 1   & 1.31       & B,V,J,H,K                      & 18 \\
NGC3893              & late-type  &   0.003226 & 18.30      & R$_{25}$                                      & 339.38     & -2.14      & 1.15       & 1   & 1.57       & B,J,H,K                        & 18 \\
NGC3938              & late-type  &   0.002699 & 12.20      & R$_{25}$                                      & 278.57     & -2.29      & 1.25       & 1   & 1.32       & fuv,nuv,B,V,R,I,J,H,K,IRAC1    & 18 \\
NGC4178              & late-type  &   0.001248 & 17.49      & R$_{25}$                                      & 326.11     & -2.45      & 1.25       & 1   & 0.95       & B,V,J,H,K                      & 18 \\
NGC4189              & late-type  &   0.007055 & 25.10      & R$_{25}$                                      & 222.06     & -2.27      & 1.18       & 1   & 1.64       & B,V,J,H,K                      & 18 \\
NGC4214              & late-type  &   0.000970 & 2.90       & 1.5$R_{25}$                                   & 59.45      & -2.74      & 0.93       & 2   & 1.03       & IRAC3.6                        & 2 \\
NGC4254              & late-type  &   0.008029 & 20.00      & R$_{25}$                                      & 720.65     & -1.88      & 1.43       & 1   & 1.59       & fuv,nuv,B,V,R,I,J,H,K,IRAC1    & 18 \\
NGC4258              & late-type  &   0.001494 & 7.30       & R$_{25}$                                      & 809.58     & -2.54      & 0.70       & 1   & 1.51       & nuv,B,V,J,H,K                  & 18 \\
NGC4294              & late-type  &   0.001184 & 17.98      & R$_{25}$                                      & 134.62     & -2.05      & 1.13       & 1   & 1.03       & U,B,V,J,H,K                    & 18 \\
NGC4303              & late-type  &   0.005224 & 17.50      & R$_{25}$                                      & 710.29     & -1.92      & 1.26       & 1   & 1.58       & U,B,V,J,H,K                    & 18 \\
NGC4321              & late-type  &   0.005240 & 20.00      & R$_{25}$                                      & 1232.35    & -2.25      & 1.21       & 1   & 1.74       & fuv,nuv,B,V,R,I,J,H,K,IRAC1    & 18 \\
NGC4394              & late-type  &   0.003075 & 16.80      & R$_{25}$                                      & 286.03     & -3.06      & 0.67       & 1   & 1.68       & U,B,V,J,H,K                    & 18 \\
NGC4402              & late-type  &   0.000774 & 15.70      & R$_{25}$                                      & 157.83     & -2.98      & 1.10       & 1   & 1.82       & nuv,B,J,H,K                    & 18 \\
NGC4449              & late-type  &   0.000690 & 4.20       & 1.5$R_{25}$                                   & 55.42      & -2.17      & 1.46       & 2   & 1.56       & IRAC3.6                        & 2 \\
NGC4501              & late-type  &   0.007609 & 10.50      & R$_{25}$                                      & 264.45     & -2.39      & 1.11       & 1   & 2.06       & U,B,V,J,H,K                    & 18 \\
NGC4519              & late-type  &   0.004056 & 27.89      & R$_{25}$                                      & 498.06     & -2.16      & 1.17       & 1   & 1.03       & U,B,V,J,H,K                    & 18 \\
NGC4535              & late-type  &   0.006551 & 17.50      & R$_{25}$                                      & 809.87     & -2.56      & 1.06       & 1   & 1.37       & B,V,J,H,K                      & 18 \\
NGC4548              & late-type  &   0.001621 & 17.50      & R$_{25}$                                      & 530.73     & -2.70      & 0.73       & 1   & 1.81       & B,V,J,H,K                      & 18 \\
NGC4561              & late-type  &   0.004693 & 12.30      & R$_{25}$                                      & 19.76      & -2.11      & 1.61       & 1   & 1.07       & U,B,V,J,H,K                    & 18 \\
NGC4569              & late-type  &  -0.000784 & 20.00      & R$_{25}$                                      & 1663.31    & -2.96      & 0.62       & 1   & 1.65       & fuv,nuv,B,V,R,I,J,H,K,IRAC1    & 18 \\
NGC4571              & late-type  &   0.001141 & 17.50      & R$_{25}$                                      & 279.34     & -2.74      & 0.88       & 1   & 1.36       & B,V,J,H,K                      & 18 \\
NGC4579              & late-type  &   0.005060 & 20.00      & R$_{25}$                                      & 693.20     & -2.50      & 0.83       & 1   & 2.26       & fuv,nuv,B,V,R,I,J,H,K,IRAC1    & 18 \\
NGC4639              & late-type  &   0.003395 & 22.28      & R$_{25}$                                      & 241.11     & -2.29      & 0.83       & 1   & 1.61       & U,B,J,H,K                      & 18 \\
NGC4647              & late-type  &   0.004700 & 17.50      & R$_{25}$                                      & 235.88     & -2.40      & 1.07       & 1   & 1.88       & U,B,J,H,K                      & 18 \\
NGC4651              & late-type  &   0.002628 & 25.95      & R$_{25}$                                      & 549.63     & -2.16      & 1.14       & 1   & 1.62       & B,V,J,H,K                      & 18 \\
NGC4654              & late-type  &   0.003489 & 17.50      & R$_{25}$                                      & 377.29     & -2.24      & 1.17       & 1   & 1.54       & U,B,V,J,H,K                    & 18 \\
NGC4689              & late-type  &   0.005390 & 17.50      & R$_{25}$                                      & 310.36     & -2.56      & 0.96       & 1   & 1.68       & B,V,J,H,K                      & 18 \\
NGC4698              & late-type  &   0.003366 & 25.10      & R$_{25}$                                      & 574.66     & -3.73      & 0.30       & 1   & 2.13       & U,B,V,J,H,K                    & 18 \\
NGC4713              & late-type  &   0.002175 & 17.05      & R$_{25}$                                      & 130.93     & -1.71      & 1.16       & 1   & 1.08       & U,B,V,J,H,K                    & 18 \\
NGC4736              & late-type  &   0.001027 & 4.70       & 1.5$R_{25}$                                   & 198.56     & -2.62      & 0.66       & 2   & 2.00       & IRAC3.6                        & 2 \\
NGC4826              & late-type  &   0.001361 & 5.60       & R$_{25}$                                      & 133.73     & -2.65      & 0.68       & 1   & 2.22       & fuv,nuv,B,V,R,I,J,H,K,IRAC1    & 18 \\
NGC5033              & late-type  &   0.002919 & 13.30      & R$_{25}$                                      & 975.99     & -2.82      & 1.01       & 1   & 1.29       & fuv,nuv,B,V,R,I,J,H,K,IRAC1    & 18 \\
NGC5055              & late-type  &   0.001614 & 10.10      & 1.5$R_{25}$                                   & 2140.08    & -3.00      & 0.92       & 2   & 1.47       & IRAC3.6                        & 2 \\
NGC5194              & late-type  &   0.001544 & 8.00       & 1.5$R_{25}$                                   & 572.56     & -2.26      & 1.00       & 2   & 1.84       & IRAC3.6                        & 2 \\
NGC5236              & late-type  &   0.001711 & 4.64       & R$_{25}$                                      & 173.57     & -1.59      & 1.71       & 1   & 2.05       & U,B,V,J,H,K                    & 18 \\
NGC5457              & late-type  &   0.000804 & 6.00       & R$_{25}$                                      & 1735.66    & -2.64      & 1.19       & 1   & 0.69       & nuv,B,V,J,H,K                  & 18 \\
NGC6207              & late-type  &   0.002842 & 20.00      & R$_{25}$                                      & 180.16     & -1.88      & 1.14       & 1   & 1.33       & B,V,J,H,K                      & 18 \\
NGC6217              & late-type  &   0.004543 & 21.40      & R$_{25}$                                      & 274.62     & -2.09      & 1.33       & 1   & 1.40       & B,V,J,H,K                      & 18 \\
NGC6503              & late-type  &   0.000083 & 5.27       & R$_{25}$                                      & 44.43      & -2.26      & 0.95       & 1   & 1.70       & B,V,J,H,K                      & 18 \\
NGC6643              & late-type  &   0.004950 & 25.80      & R$_{25}$                                      & 512.69     & -1.99      & 1.19       & 1   & 1.55       & B,V,J,H,K                      & 18 \\
NGC6946              & late-type  &   0.000133 & 5.90       & 1.5$R_{25}$                                   & 678.87     & -2.32      & 1.18       & 2   & 1.67       & IRAC3.6                        & 2 \\
NGC7331              & late-type  &   0.002722 & 14.70      & 1.5$R_{25}$                                   & 2715.47    & -2.96      & 0.81       & 2   & 1.47       & IRAC3.6                        & 2 \\
NGC7793              & late-type  &   0.000757 & 3.90       & 1.5$R_{25}$                                   & 254.47     & -3.03      & 0.69       & 2   & 1.09       & IRAC3.6                        & 2 \\
IC1141               & early-type &   0.014670 & 68.00      & $R_{\rm max}$[CO]($\sim$0.4$R_{25}$)          & 19.26      & -1.11      & 1.74       & 3,18 & 2.89       & u,g,r,i,z,J,H,K                & 18 \\
NGC2320              & early-type &   0.019827 & 83.00      & $R_{\rm max}$[CO]($\sim$-0.2$R_{25}$)         & 50.27      & -1.11      & 1.90       & 4   & 3.07       & J,H,K                          & 18 \\
NGC2768              & early-type &   0.004580 & 21.80      & $R_{\rm max}$[CO]($\sim$-0.1$R_{25}$)         & 1.13       & -1.66      & 1.66       & 4   & 4.06       & u,g,r,i,z,J,H,K                & 18 \\
NGC3032              & early-type &   0.005114 & 25.20      & $R_{\rm max}$[CO]($\sim$0.3$R_{25}$)          & 5.28       & -1.32      & 1.82       & 3,12 & 2.68       & u,g,r,i,z,J,H,K                & 18 \\
NGC3073              & early-type &   0.003853 & 21.10      & $R_{\rm max}$[CO]($\sim$0.3$R_{25}$)          & 3.70       & -2.40      & 0.66       & 3,13 & 2.33       & u,g,r,i,z,J,H,K                & 18 \\
NGC3489              & early-type &   0.002258 & 11.80      & $R_{\rm max}$[CO]($\sim$-0.1$R_{25}$)         & 2.01       & -1.86      & 0.97       & 4   & 3.62       & u,g,r,i,z,J,H,K                & 18 \\
NGC3773              & early-type &   0.003276 & 10.50      & $R_{\rm max}$[CO]($\sim$0.3$R_{25}$)          & 0.92       & -1.24      & 1.17       & 3,14 & 1.91       & u,g,r,i,z,J,H,K                & 18 \\
NGC3870              & early-type &   0.002522 & 14.50      & $R_{\rm max}$[CO]($\sim$0.4$R_{25}$)          & 2.28       & -1.42      & 1.07       & 3,15 & 2.08       & u,g,r,i,z,J,H,K                & 18 \\
NGC4150              & early-type &   0.000754 & 13.40      & $R_{\rm max}$[CO]($\sim$-0.2$R_{25}$)         & 1.13       & -1.62      & 1.67       & 4   & 3.40       & u,g,r,i,z,J,H,K                & 18 \\
NGC4459              & early-type &   0.004036 & 16.10      & $R_{\rm max}$[CO]($\sim$-0.1$R_{25}$)         & 3.14       & -1.67      & 1.68       & 4   & 3.79       & u,g,r,i,z,J,H,K                & 18 \\
NGC4477              & early-type &   0.004520 & 16.50      & $R_{\rm max}$[CO]($\sim$-0.1$R_{25}$)         & 0.50       & -1.29      & 1.62       & 4   & 4.16       & u,g,r,i,z,J,H,K                & 18 \\
NGC4526              & early-type &   0.001494 & 16.40      & $R_{\rm max}$[CO]($\sim$-0.0$R_{25}$)         & 3.14       & -1.16      & 2.19       & 4   & 3.83       & u,g,r,i,z,J,H,K                & 18 \\
NGC4550              & early-type &   0.001271 & 15.50      & $R_{\rm max}$[CO]($\sim$-0.1$R_{25}$)         & 0.50       & -1.92      & 1.20       & 4   & 3.79       & u,g,r,i,z,J,H,K                & 18 \\
NGC5173              & early-type &   0.008069 & 41.20      & $R_{\rm max}$[CO]($\sim$0.3$R_{25}$)          & 14.12      & -2.22      & 1.04       & 3,16 & 2.99       & u,g,r,i,z,J,H,K                & 18 \\
NGC524               & early-type &   0.007935 & 23.30      & $R_{\rm max}$[CO]($\sim$-0.1$R_{25}$)         & 3.80       & -2.04      & 1.19       & 4   & 3.46       & J,H,K                          & 18 \\
NGC5338              & early-type &   0.002722 & 10.30      & $R_{\rm max}$[CO]($\sim$0.1$R_{25}$)          & 0.44       & -1.96      & 1.28       & 3,18 & 2.49       & u,g,r,i,z,J,H,K                & 18 \\
NGC5666              & early-type &   0.007408 & 35.00      & $R_{\rm max}$[CO]($\sim$-0.4$R_{25}$)         & 22.90      & -1.08      & 1.34       & 4   & 2.44       & u,g,r,i,z,J,H,K                & 18 \\
UGC6805              & early-type &   0.003864 & 20.30      & $R_{\rm max}$[CO]($\sim$0.4$R_{25}$)          & 2.47       & -1.82      & 1.02       & 3,18 & 2.31       & u,g,r,i,z,J,H,K                & 18 \\
UGC9562              & early-type &   0.004310 & 25.20      & $R_{\rm max}$[CO]($\sim$0.4$R_{25}$)          & 6.88       & -2.52      & 0.20       & 3,17 & 1.77       & u,g,r,i,z,J,H,K                & 18 \\
DDO154               & LSB        &   0.001248 & 4.30       & 1.5$R_{25}$                                   & 10.18      & -3.31      & 1.70       & 2   & 0.09       & IRAC3.6                        & 2 \\
HOI                  & LSB        &   0.000465 & 3.80       & 1.5$R_{25}$                                   & 22.90      & -3.41      & 0.97       & 2   & 0.04       & IRAC3.6                        & 2 \\
HOII                 & LSB        &   0.000474 & 3.40       & 1.5$R_{25}$                                   & 96.77      & -3.30      & 0.94       & 2   & 0.31       & IRAC3.6                        & 2 \\
IC2574               & LSB        &   0.000190 & 4.00       & 1.5$R_{25}$                                   & 397.61     & -3.75      & 0.72       & 2   & 0.10       & IRAC3.6                        & 2 \\
LSBCF561-01          & LSB        &   0.016034 & 67.00      & $R_{\rm max}$[HI,UV]($\sim$1.6$R_{25}$)       & 338.84     & -3.39      & 0.75       & 5   & 0.42       & u,g,r,i,z                      & 18 \\
LSBCF563-01          & LSB        &   0.011681 & 49.00      & $R_{\rm max}$[HI,UV]($\sim$3.5$R_{25}$)       & 1071.52    & -3.97      & 0.70       & 5   & -0.21      & u,g,r,i,z                      & 18 \\
LSBCF563-V01         & LSB        &   0.012976 & 54.00      & $R_{\rm max}$[HI,UV]($\sim$2.3$R_{25}$)       & 186.21     & -3.98      & 0.64       & 5   & 0.24       & u,g,r,i,z                      & 18 \\
LSBCF564-V03         & LSB        &   0.001612 & 9.00       & $R_{\rm max}$[HI,UV]($\sim$3.6$R_{25}$)       & 7.59       & -3.97      & 0.45       & 5   & 0.14       & u,g,r,i,z                      & 18 \\
LSBCF565-V02         & LSB        &   0.012278 & 51.00      & $R_{\rm max}$[HI,UV]($\sim$3.6$R_{25}$)       & 239.88     & -3.98      & 0.52       & 5   & -0.14      & u,g,r,i,z                      & 18 \\
LSBCF568-01          & LSB        &   0.021762 & 91.00      & $R_{\rm max}$[HI,UV]($\sim$2.0$R_{25}$)       & 741.31     & -3.60      & 0.99       & 5   & 0.27       & u,g,r,i,z                      & 18 \\
LSBCF568-03          & LSB        &   0.019717 & 83.00      & $R_{\rm max}$[HI,UV]($\sim$2.1$R_{25}$)       & 891.25     & -3.59      & 0.76       & 5   & 0.41       & u,g,r,i,z                      & 18 \\
LSBCF568-06          & LSB        &   0.046132 & 205.00     & $R_{\rm max}$[HI,UV]                          & 16595.86   & -3.92      & 0.41       & 5   & 1.14       & u,g,r,i,z                      & 18 \\
LSBCF568-V01         & LSB        &   0.019243 & 86.00      & $R_{\rm max}$[HI,UV]($\sim$2.5$R_{25}$)       & 794.33     & -3.62      & 0.71       & 5   & 0.14       & u,g,r,i,z                      & 18 \\
LSBCF574-01          & LSB        &   0.022979 & 103.00     & $R_{\rm max}$[HI,UV]($\sim$0.9$R_{25}$)       & 758.58     & -3.54      & 0.57       & 5   & 0.42       & u,g,r,i,z                      & 18 \\
LSBCF574-02          & LSB        &   0.021081 & 94.00      & $R_{\rm max}$[HI,UV]($\sim$2.8$R_{25}$)       & 380.19     & -3.51      & 0.62       & 5   & 0.12       & u,g,r,i,z                      & 18 \\
LSBCF577-V01         & LSB        &   0.025978 & 114.00     & $R_{\rm max}$[HI,UV]($\sim$1.3$R_{25}$)       & 436.52     & -3.18      & 0.50       & 5   & 0.12       & u,g,r,i,z                      & 18 \\
LSBCF579-V01         & LSB        &   0.020995 & 91.00      & $R_{\rm max}$[HI,UV]($\sim$1.6$R_{25}$)       & 977.24     & -3.74      & 0.63       & 5   & 0.66       & u,g,r,i,z                      & 18 \\
LSBCF583-01          & LSB        &   0.007552 & 34.00      & $R_{\rm max}$[HI,UV]($\sim$3.7$R_{25}$)       & 501.19     & -3.94      & 0.76       & 5   & -0.57      & u,g,r,i,z                      & 18 \\
Malin1               & LSB        &   0.082557 & 380.00     & $R_{\rm max}$[HI,UV]($\sim$11.3$R_{25}$)      & 37153.54   & -4.39      & 0.48       & 5   & 0.36       & u,g,r,i,z,J,H,K,IRAC36         & 18 \\
NGC0925              & LSB        &   0.001845 & 9.20       & 1.5$R_{25}$                                   & 1425.31    & -3.40      & 0.66       & 2   & 0.75       & IRAC3.6                        & 2 \\
UGC5750              & LSB        &   0.013893 & 60.00      & $R_{\rm max}$[HI,UV]($\sim$2.0$R_{25}$)       & 870.96     & -3.69      & 0.56       & 5   & -0.32      & u,g,r,i,z                      & 18 \\
UGC5999              & LSB        &   0.011314 & 49.00      & $R_{\rm max}$[HI,UV]($\sim$1.9$R_{25}$)       & 707.95     & -3.57      & 0.80       & 5   & -0.03      & u,g,r,i,z                      & 18 \\
UGC6614              & LSB        &   0.021188 & 92.00      & $R_{\rm max}$[HI,UV]($\sim$3.1$R_{25}$)       & 9120.11    & -3.85      & 0.55       & 5   & 0.84       & u,g,r,i,z,J,H,K,IRAC36         & 18 \\
Arp220               & z=0-LIRGs  &   0.018126 & 71.00      & $R_{\rm max}$[CO]($\sim$0.1$R_{25}$)          & 7.07       & 1.39       & 3.41       & 6   & 2.89       & B,V,R,J,H,K                    & 18 \\
IRAS10190+1322       & z=0-LIRGs  &   0.076600 & 340.00     & $R_{\rm max}$[CO]($\sim$0.2$R_{25}$)          & 45.36      & 0.40       & 2.97       & 7   & 2.91       & J,H,K                          & 18 \\
Mrk273               & z=0-LIRGs  &   0.037780 & 151.00     & $R_{\rm max}$[CO]($\sim$0.2$R_{25}$)          & 20.43      & 0.78       & 3.19       & 8   & 2.87       & u,g,r,i,z,J,H,K                & 18 \\
NGC6090              & z=0-LIRGs  &   0.029304 & 123.30     & $R_{\rm max}$[CO]($\sim$1.2$R_{25}$)          & 268.80     & -0.87      & 2.00       & 9   & 2.02       & B,V,I,J,H,K                    & 18 \\
NGC6240              & z=0-LIRGs  &   0.024480 & 100.90     & $R_{\rm max}$[CO]($\sim$0.2$R_{25}$)          & 147.41     & -0.29      & 2.36       & 9   & 2.86       & U,B,V,J,H,K                    & 18 \\
NGC7674              & z=0-LIRGs  &   0.028924 & 115.30     & $R_{\rm max}$[CO]($\sim$0.3$R_{25}$)          & 118.82     & -0.51      & 2.00       & 9   & 2.49       & B,V,J,H,K                      & 18 \\
VV114                & z=0-LIRGs  &   0.020067 & 80.00      & $R_{\rm max}$[CO]($\sim$0.2$R_{25}$)          & 27.34      & 0.24       & 3.23       & 10   & 2.39       & Jb,J,H,K                       & 18 \\
BzK16000             & high-z-SFG &       1.52 & 11089      & $R_{1/2}$[UV/optical, CO]                     & 52.81      & 0.16       & 2.79       & 11   & 2.94       & U,B,V,R,I,z,J,H,K              & 18 \\
BzK17999             & high-z-SFG &       1.41 & 10107      & $R_{1/2}$[UV/optical, CO]                     & 40.72      & 0.26       & 2.97       & 11   & 3.15       & U,B,V,R,I,z,J,H,K              & 18 \\
BzK210000            & high-z-SFG &       1.52 & 11089      & $R_{1/2}$[UV/optical, CO]                     & 58.09      & 0.28       & 2.93       & 11   & 2.88       & U,B,V,R,I,z,J,H,K              & 18 \\
BzK4171              & high-z-SFG &       1.47 & 10640      & $R_{1/2}$[UV/optical, CO]                     & 43.01      & 0.08       & 3.03       & 11   & 3.01       & U,B,V,R,I,z,J,H,K              & 18 \\
EGS12007881          & high-z-SFG &       1.17 & 8024       & $R_{1/2}$[UV/optical]                         & 237.79     & -0.72      & 2.24       & 11   & 2.09       & B,R,I,Ks                       & 18 \\
EGS12011767          & high-z-SFG &       1.28 & 8967       & $R_{1/2}$[UV/optical]                         & 167.42     & -0.85      & 1.89       & 11   & 2.20       & B,R,I,Ks                       & 18 \\
EGS13003805          & high-z-SFG &       1.23 & 8536       & $R_{1/2}$[UV/optical]                         & 84.95      & -0.12      & 3.02       & 11   & 2.94       & B,R,I,Ks                       & 18 \\
EGS13004291          & high-z-SFG &       1.20 & 8279       & $R_{1/2}$[UV/optical]                         & 162.86     & -0.28      & 2.93       & 11   & 2.85       & B,R,I,Ks                       & 18 \\
EGS13004661          & high-z-SFG &       1.19 & 8194       & $R_{1/2}$[UV/optical]                         & 149.57     & -0.56      & 1.90       & 11   & 2.26       & B,R,I,Ks                       & 18 \\
EGS13011148          & high-z-SFG &       1.17 & 8024       & $R_{1/2}$[UV/optical]                         & 58.09      & -0.45      & 2.41       & 11   & 2.87       & B,R,I,Ks                       & 18 \\
EGS13011439          & high-z-SFG &       1.10 & 7434       & $R_{1/2}$[UV/optical]                         & 15.21      & 0.47       & 3.17       & 11   & 3.46       & B,R,I,Ks                       & 18 \\
EGS13017614          & high-z-SFG &       1.18 & 8109       & $R_{1/2}$[UV/optical]                         & 124.69     & -0.53      & 2.57       & 11   & 2.76       & B,R,I,Ks                       & 18 \\
EGS13035123          & high-z-SFG &       1.12 & 7602       & $R_{1/2}$[UV/optical]                         & 232.35     & -0.57      & 2.44       & 11   & 2.86       & B,R,I,Ks                       & 18 \\
Q1623-BX599          & high-z-SFG &       2.33 & 18723      & $R_{1/2}$[H$\alpha$]                          & 24.63      & 0.42       & 3.51       & 11   & 2.64       & U,G,R,J,Ks                     & 18 \\
Q1700-BX691          & high-z-SFG &       2.19 & 17362      & $R_{1/2}$[H$\alpha$]                          & 141.03     & -0.73      & 2.09       & 11   & 1.97       & U,G,R,J,Ks                     & 18 \\
Q1700-MD174          & high-z-SFG &       2.34 & 18821      & $R_{1/2}$[H$\alpha$]                          & 40.72      & 0.15       & 3.29       & 11   & 2.91       & U,G,R,J,Ks                     & 18 \\
Q1700-MD69           & high-z-SFG &       2.29 & 18333      & $R_{1/2}$[H$\alpha$]                          & 277.59     & -0.54      & 2.29       & 11   & 1.87       & U,G,R,J,Ks                     & 18 \\
Q1700-MD94           & high-z-SFG &       2.34 & 18821      & $R_{1/2}$[H$\alpha$]                          & 289.53     & -0.25      & 2.96       & 11   & 2.32       & U,G,R,J,Ks                     & 18 \\
Q2343-BX442          & high-z-SFG &       2.18 & 17265      & $R_{1/2}$[H$\alpha$]                          & 141.03     & -0.47      & 2.55       & 11   & 2.16       & U,G,R,J,Ks                     & 18 \\
Q2343-BX610          & high-z-SFG &       2.21 & 17555      & $R_{1/2}$[H$\alpha$]                          & 45.36      & 0.36       & 3.40       & 11   & 2.88       & U,G,R,J,Ks                     & 18 \\
Q2343-MD59           & high-z-SFG &       2.01 & 15634      & $R_{1/2}$[H$\alpha$]                          & 95.03      & -0.68      & 2.68       & 11   & 1.93       & U,G,R,J,Ks                     & 18 \\
SMMJ123549+6215      & high-z-SMG &       2.20 & 17458      & $R_{1/2}$[CO]                                 & 2.54       & 2.25       & 4.65       & 11   & 4.24       & U,B,V,R,I,z,J                  & 18 \\
SMMJ123634+6212      & high-z-SMG &       1.22 & 8450       & $R_{1/2}$[CO]                                 & 52.81      & 0.64       & 3.19       & 11   & 2.78       & U,B,V,R,I,z,J,K,IRAC36         & 18 \\
SMMJ123707+6214      & high-z-SMG &       2.49 & 20297      & $R_{1/2}$[CO]                                 & 24.63      & 1.01       & 3.34       & 11   & 3.50       & U,B,V,R,I,z,J,K,IRAC36         & 18 \\
SMMJ131201+4242      & high-z-SMG &       3.41 & 29645      & $R_{1/2}$[CO]                                 & 28.27      & 1.07       & 3.57       & 11   & 3.06       & B,R,I,z,J,K,IRAC36             & 18 \\
SMMJ131232+4239      & high-z-SMG &       2.33 & 18723      & $R_{1/2}$[CO]                                 & 12.57      & 1.31       & 3.80       & 11   & 3.74       & B,R,I,z,J,K,IRAC36             & 18 \\
SMMJ163650+4057      & high-z-SMG &       2.39 & 19311      & $R_{1/2}$[CO]                                 & 18.10      & 1.39       & 4.02       & 11   & 3.65       & B,V,R,I,K,IRAC36               & 18 \\
SMMJ163658+4105      & high-z-SMG &       2.45 & 19902      & $R_{1/2}$[CO]                                 & 2.01       & 2.45       & 4.90       & 11   & 4.61       & B,R,I,K,IRAC36                 & 18 \\
\enddata
\tablecomments{ Col.(1): Name. Col.(2): The galaxy type in this paper.
Col.(3): The redshift. Col.(3): The distance in Mpc that is used in 
this work.  Col.(4): The definition of the aperture
used to  calculate the surface density  of SFR, gas  and stellar mass.
R$_{25}$ means  the isophotal radius at  25 mag/arcsec$^{-2}$ (usually
in B band). $R_{\rm max}$ gives  the maximum extent of the galaxy at a
given  wavelength that  is  indicated in  square  brackets. Its  value
relative   to    R$_{25}$   is   listed    in   parentheses   whenever
possible. $R_{1/2}$  is the  half light radius  at a  given wavelength
that is indicated in  square brackets.  Col.(5): The de-projected area
used  to   derive  surface   densities.   Col.(6):  The   SFR  surface
density. Col.(7): The gas surface density. Col.(8): The references and
therein  for the  SFR  and  gas surface  density  data. For  those with two
references, the  first one is for  H$_{2}$ and SFR  data while the
second   is for  HI data. Col.(9):  The stellar  mass surface
density.  Col.(10):  The wavelength band  used to measure  the stellar
mass.  Col.(11):  The references  for  the stellar  mass data.\\
References:    
1-\citet{Kennicutt98a}, 2-\citet{Leroy08}, 3-\citet{Wei10}, 4-\citet{Crocker11}, 5-\citet{Wyder09}, 6-\citet{Scoville97}, 7-\citet{Gracia-Carpio07}, 8-\citet{Yun95}, 9-\citet{Bryant99}, 10-\citet{Yun94}, 11-\citet{Genzel10}, 12-\citet{Oosterloo10}, 13-\citet{Irwin87}, 14-\citet{vanDriel91}, 15-\citet{Sage06}, 16-\citet{Knapp84}, 17-\citet{Cox01}, 18-this-work. }
\end{deluxetable}
\clearpage
\end{landscape}

\begin{deluxetable}{ll}
\tabletypesize{\scriptsize}
\tablecolumns{7}
\tablecaption{\label{BC_model} The Parameters of Stellar Synthesis Models.}
\tablewidth{0pt}
\tablehead{ 
\colhead{parameters}           &     value  \\
}
\startdata
simple stellar populations                                    &  \citet{Chabrier03} IMF and Padova 1994 evolutionary tracks\\
metallicity                                                   &  0.0004, 0.004, 0.008, 0.02 ($Z_{\odot}$), 0.05 \\
visual extinction $\tau_{v}$                                   &  [0.0, 10.0] with a step of 1 \\
fraction of $\tau_{V}$ arising from the ambient ISM            &  0.3 \\
fraction of ejected gas to be recycled in stars               &  0.0 \\
star formation history (SFH)                                  &  exponential decline\\
e-folding time $\tau$ for   exponential SFH                   &  [0.03, 22.4] Gyr with a step of 0.15 in logarithm, plus 100 Gyr       \\
galaxy age                                                    &  [0.05, 12.6] Gyr with a step of 0.12 in logarithm \\
\enddata
\end{deluxetable}


\begin{deluxetable}{llllllllllllllll}
\tabletypesize{\scriptsize}
\tablecolumns{15}
\tablecaption{\label{samp_ind_region} The Sample Of Spiral Galaxies}
\tablewidth{0pt}
\tablehead{ 
\colhead{Name}  &  \colhead{Position (J2000)}  &  \colhead{Type}  & \colhead{Dist}      & \colhead{$R_{25}$}   & \colhead{$i$}      & \colhead{PA}  \\
\colhead{ }     &  \colhead{}                  & \colhead{}       & \colhead{[Mpc]}     & \colhead{[arcsec]} & \colhead{[$\circ$]} & \colhead{[$\circ$]}    \\
\colhead{(1)}   &  \colhead{(2)}               & \colhead{(3)}    & \colhead{(4)}       & \colhead{(5)}       & \colhead{(6)}   & \colhead{(7)}    \\
}
\startdata
NGC0628              &  01 36 41.8  +15 47 00 & SAc        & 7.30       & 293.17     & 7.00       & 20.00   \\  
NGC2841              &  09 22 02.6  +50 58 35 & SAb        & 14.10      & 207.55     & 74.00      & 153.00  \\  
NGC3184              &  10 18 17.0  +41 25 28 & SABcd      & 11.10      & 222.39     & 16.00      & 179.00  \\  
NGC3351              &  10 43 57.7  +11 42 14 & SBb        & 10.10      & 217.33     & 41.00      & 192.00  \\  
NGC3521              &  11 05 48.6  -00 02 09 & SABbc      & 10.70      & 249.53     & 73.00      & 340.00  \\  
NGC3627              &  11 20 15.0  +12 59 30 & SABb       & 9.30       & 306.99     & 62.00      & 173.00  \\  
NGC4736              &  12 50 53.0  +41 07 13 & SAab       & 4.70       & 232.87     & 41.00      & 296.00  \\  
NGC4826              &  12 56 43.6  +21 41 00 & SAab       & 7.50       & 314.14     & 65.00      & 121.00  \\  
NGC5055              &  13 15 49.2  +42 01 45 & SAbc       & 10.10      & 352.47     & 59.00      & 102.00  \\  
NGC5194              &  13 29 52.7  +47 11 43 & SABbc      & 8.00       & 232.87     & 42.00      & 172.00  \\  
NGC6946              &  20 34 52.2  +60 09 14 & SABcd      & 5.90       & 344.45     & 33.00      & 243.00  \\  
NGC7331              &  22 37 04.1  +34 24 57 & SAb        & 14.70      & 273.60     & 76.00      & 168.00  \\  
\enddata
\tablecomments{Col.(1):  Name.  Col.(2):  The position.  Col.(3):  The
Hubble  type.  Col.(4):  The   distance.  Col.(5):  The  optical  size
$R_{25}$. Col.(6): The inclination.  Col.(7): The position angle.}
\end{deluxetable}

\begin{deluxetable}{llllllllllllllll}
\tabletypesize{\scriptsize}
\tablecolumns{15}
\tablecaption{\label{fit_result} The Parameters of the Best Fit}
\tablewidth{0pt}
\tablehead{ 
\colhead{Figures}  &  \colhead{Correlations}    &  \colhead{$a$}  & \colhead{$b$}      & \colhead{$\delta$}  & \colhead{$\rho$}  & \colhead{$\sigma$}  \\
\colhead{(1)}      &  \colhead{(2)}          & \colhead{(3)}          & \colhead{(4)}           & \colhead{(5)}    & \colhead{(6)}   & \colhead{(7)}    \\
}
\startdata
Fig.~\ref{SFE_SBMstar}(a)   &   SFE-$\Sigma_{\rm star}$                                                                 & -10.28$\pm$0.08 & 0.48$\pm$0.04 & 0.123 & 0.97$\pm$0.03 & 0.41  \\
Fig.~\ref{SFE_SBMstar}(b)   &   SFE-$\Sigma_{\rm gas}$                                                                  &  -9.85$\pm$0.08 & 0.35$\pm$0.04 & 0.092 & 0.97$\pm$0.03 & 0.49  \\
Fig.~\ref{SFR_MgasMstar}(a) & $\Sigma_{\rm SFR}$-$\Sigma_{\rm star}^{0.5}\Sigma_{\rm gas}$                                   &  -4.40$\pm$0.08 & 1.03$\pm$0.03 & 0.131 & 1.00$\pm$0.00 & 0.42  \\
Fig.~\ref{SFR_MgasMstar}(b) &    KS Law                                                                               &  -3.90$\pm$0.11 & 1.38$\pm$0.06 & 0.112 & 1.00$\pm$0.00 & 0.49  \\
 
Fig.~\ref{result_diff_a}(a)  &   SFE-$\Sigma_{\rm star}$ (diff. $\alpha$)                                               & -10.35$\pm$0.09 & 0.55$\pm$0.04 & 0.171 & 0.96$\pm$0.03 & 0.47  \\
Fig.~\ref{result_diff_a}(b)  &   SFE-$\Sigma_{\rm gas}$ (diff. $\alpha$)                                                &  -9.97$\pm$0.10 & 0.49$\pm$0.06 & 0.172 & 0.93$\pm$0.05 & 0.54  \\
Fig.~\ref{result_diff_a}(c)  &    $\Sigma_{\rm SFR}$-$\Sigma_{\rm star}^{0.5}\Sigma_{\rm gas}$ (diff. $\alpha$)              &  -4.51$\pm$0.09 & 1.10$\pm$0.03 & 0.180 & 0.99$\pm$0.01 & 0.47  \\
Fig.~\ref{result_diff_a}(d)  &    KS Law (diff. $\alpha$)                                                             &  -4.17$\pm$0.14 & 1.65$\pm$0.09 & 0.125 & 1.00$\pm$0.00 & 0.55  \\
\enddata
\tablecomments{Col.(1):   The   corresponding   figure  to   a   given
correlation.  Note that  the LSB  and early-type  galaxies  are always
excluded for  the KS law  or the equivalent  version (SFE-$\Sigma_{\rm
gas}$).  ``diff.  $\alpha$'' means different  CO-to-H$_{2}$ conversion
factors  for   mergers  and  non-mergers.   Col.(2):  The   Y  vs.   X
correlations.  Col.(3)-Col.(5):  log(Y) = $a$  + $b$*log(X)+$\epsilon$
as  given by linmix$\_$err.pro  that accounts  for measured  errors on
both  X-  and  Y-axis   \citep{Kelly07},  where  ($a$,  $b$)  are  the
regression coefficients and $\epsilon$ is the intrinsic random scatter
about  the  regression  and  has  a  mean  of  zero  and  variance  of
$\delta^{2}$.  Col.(6):  The linear correlation  efficiency.  Col.(7):
The observed standard deviation of all data points from the best fit.}
\end{deluxetable}




\begin{thebibliography}{}

\bibitem[Bigiel et al.(2008)]{Bigiel08} Bigiel, F., Leroy, A., Walter, F., Brinks, E., de Blok, W.~J.~G., Madore, B., \& Thornley, M.~D.\ 2008, \aj, 136, 2846 


\bibitem[Blanton et al.(2003)]{Blanton03} Blanton, M.~R., et al.\ 2003, \apj, 594, 186 



\bibitem[Blitz \& Rosolowsky(2004)]{Blitz04} Blitz, L., \& Rosolowsky, E.\ 2004, \apjl, 612, L29 

\bibitem[Blitz \& Rosolowsky(2006)]{Blitz06} Blitz, L., \& Rosolowsky, E.\ 2006, \apj, 650, 933 

\bibitem[Boissier et al.(2003)]{Boissier03} Boissier, S., Prantzos, N., Boselli, A., \& Gavazzi, G.\ 2003, \mnras, 346, 1215 





\bibitem[Bouch{\'e} et al.(2010)]{Bouche10} Bouch{\'e}, N., et al.\ 2010, \apj, 718, 1001 

\bibitem[Bouwens et al.(2009)]{Bouwens09} Bouwens, R.~J., et al.\ 2009, \apj, 705, 936 

\bibitem[Brinchmann et al.(2004)]{Brinchmann04} Brinchmann, J., Charlot, S., White, S.~D.~M., Tremonti, C., Kauffmann, G., Heckman, T., \& Brinkmann, J.\ 2004, \mnras, 351, 1151 

\bibitem[Brosch et al.(1998)]{Brosch98} Brosch, N., Heller, A., \& Almoznino, E.\ 1998, \apj, 504, 720 

\bibitem[Bruzual \& Charlot(2003)]{BC03} Bruzual, G., \& Charlot, S.\ 2003, \mnras, 344, 1000 

\bibitem[Bryant \& Scoville(1999)]{Bryant99} Bryant, P.~M., \& Scoville, N.~Z.\ 1999, \aj, 117, 2632 

\bibitem[Bruzual (2007)]{Bruzual07} Bruzual A, G.\ 2007, arXiv:astro-ph/0703052 


\bibitem[Chabrier(2003)]{Chabrier03} Chabrier, G.\ 2003, \pasp, 115, 763 

\bibitem[Chen et al.(2009)]{Chen09} Chen, Y.-M., Wild, V., Kauffmann, G., Blaizot, J., Davis, M., Noeske, K., Wang, J.-M., \& Willmer, C.\ 2009, \mnras, 393, 406 

\bibitem[Crosthwaite \& Turner(2007)]{Crosthwaite07} Crosthwaite, L.~P., \& Turner, J.~L.\ 2007, \aj, 134, 1827 

\bibitem[Courteau(1996)]{Courteau96} Courteau, S.\ 1996, \apjs, 103, 363 

\bibitem[Cox et al.(2001)]{Cox01} Cox, A.~L., Sparke, L.~S., Watson, A.~M., \& van Moorsel, G.\ 2001, \aj, 121, 692 

\bibitem[Crocker et al.(2011)]{Crocker11} Crocker, A.~F., Bureau, M., Young, L.~M., \& Combes, F.\ 2011, \mnras, 410, 1197 

\bibitem[Daddi et al.(2007)]{Daddi07} Daddi, E., et al.\ 2007, \apj, 670, 156 


\bibitem[Daddi et al.(2010)]{Daddi10a} Daddi, E., et al.\ 2010, \apj, 713, 686 

\bibitem[Daddi et al.(2010b)]{Daddi10b} Daddi, E., et al.\ 2010, \apj, 713, 686 



\bibitem[Dav{\'e}(2008)]{Dave08} Dav{\'e}, R.\ 2008, \mnras, 385, 147 

\bibitem[de Vaucouleurs et al.(1976)]{deVaucouleurs76} de Vaucouleurs, 
G., de Vaucouleurs, A., \& Corwin, H.~G.\ 1976, University of Texas Monographs in Astronomy, Austin: University of Texas Press, 1976,  

\bibitem[Dickman et al.(1986)]{Dickman86} Dickman, R.~L., Snell, 
R.~L., \& Schloerb, F.~P.\ 1986, \apj, 309, 326 



\bibitem[Downes \& Solomon(1998)]{Downes98} Downes, D., \& Solomon, P.~M.\ 1998, \apj, 507, 615 

\bibitem[Dunne et al.(2009)]{Dunne09} Dunne, L., et al.\ 2009, \mnras, 394, 3 

\bibitem[Dutton et al.(2010)]{Dutton10} Dutton, A.~A., van den Bosch, F.~C., \& Dekel, A.\ 2010, \mnras, 405, 1690 


\bibitem[Elbaz et al.(2007)]{Elbaz07} Elbaz, D., et al.\ 2007, \aap, 468, 33 


\bibitem[Elmegreen(1993)]{Elmegreen93} Elmegreen, B.~G.\ 1993, \apj, 411, 170 


\bibitem[Elmegreen \& Parravano(1994)]{Elmegreen94} Elmegreen, B.~G., \& Parravano, A.\ 1994, \apjl, 435, L121 


\bibitem[Elmegreen(1997)]{Elmegreen97} Elmegreen, B.~G.\ 1997, \apj, 486, 944 

\bibitem[Erb et al.(2006)]{Erb06} Erb, D.~K., Steidel, C.~C., Shapley, A.~E., Pettini, M., Reddy, N.~A., \& Adelberger, K.~L.\ 2006, \apj, 646, 107 




\bibitem[Fry et al.(1999)]{Fry99} Fry, A.~M., Morrison, H.~L., Harding, P., \& Boroson, T.~A.\ 1999, \aj, 118, 1209 

\bibitem[Gao \& Solomon(2004)]{Gao04} Gao, Y., \& Solomon, P.~M.\ 2004, \apj, 606, 271 


\bibitem[Genzel et al.(2010)]{Genzel10} Genzel, R., et al.\ 2010, arXiv:1003.5180 

\bibitem[Gil de Paz et al.(2007)]{GildePaz07} Gil de Paz, A., et al.\ 2007, \apjs, 173, 185 






\bibitem[Hainline et al.(2010)]{Hainline10} Hainline, L.~J., Blain, A.~W., Smail, I., Alexander, D.~M., Armus, L., Chapman, S.~C., \& Ivison, R.~J.\ 2010, arXiv:1006.0238 

\bibitem[Helfer et al.(2003)]{Helfer03} Helfer, T.~T., Thornley, M.~D., Regan, M.~W., Wong, T., Sheth, K., Vogel, S.~N., Blitz, L., \& Bock, D.~C.-J.\ 2003, \apjs, 145, 259 


\bibitem[Hinz et al.(2007)]{Hinz07} Hinz, J.~L., Rieke, M.~J., Rieke, G.~H., Willmer, C.~N.~A., Misselt, K., Engelbracht, C.~W., Blaylock, M., \& Pickering, T.~E.\ 2007, \apj, 663, 895

\bibitem[Ho et al.(1997)]{Ho97} Ho, L.~C., Filippenko, A.~V., \& Sargent, W.~L.~W.\ 1997, \apj, 487, 591 

\bibitem[Hollenbach \& Tielens(1999)]{Hollenbach99} Hollenbach, D.~J., \& Tielens, A.~G.~G.~M.\ 1999, Reviews of Modern Physics, 71, 173 

\bibitem[Hopkins et al.(2009a)]{Hopkins09a} Hopkins, P.~F., et al.\ 2009a, \mnras, 397, 802 

\bibitem[Hopkins et al.(2009b)]{Hopkins09b} Hopkins, P.~F., Cox, T.~J., Younger, J.~D., \& Hernquist, L.\ 2009b, \apj, 691, 1168 

\bibitem[Hunter et al.(1998)]{Hunter98} Hunter, D.~A., Elmegreen, B.~G., \& Baker, A.~L.\ 1998, \apj, 493, 595 

\bibitem[Hunter \& Elmegreen(2004)]{Hunter04} Hunter, D.~A., \& Elmegreen, B.~G.\ 2004, \aj, 128, 2170 

\bibitem[Irwin et al.(1987)]{Irwin87} Irwin, J.~A., Seaquist, E.~R., Taylor, A.~R., \& Duric, N.\ 1987, \apjl, 313, L91 

\bibitem[Isobe et al.(1990)]{Isobe90} Isobe, T., Feigelson, E.~D., Akritas, M.~G., \& Babu, G.~J.\ 1990, \apj, 364, 104 

\bibitem[Jogee et al.(2005)]{Jogee05} Jogee, S., Scoville, N., \& Kenney, J.~D.~P.\ 2005, \apj, 630, 837 

\bibitem[Karim et al.(2011)]{Karim11} Karim, A., Schinnerer, E., VLA-COSMOS, t., \& COSMOS collaborations 2011, arXiv:1102.1423 


\bibitem[Kelly(2007)]{Kelly07} Kelly, B.~C.\ 2007, \apj, 665, 1489 

\bibitem[Kennicutt(1989)]{Kennicutt89} Kennicutt, R.~C., Jr.\ 1989, \apj, 344, 685 

\bibitem[Kennicutt(1998a)]{Kennicutt98a} Kennicutt, R.~C., Jr.\ 1998a, \apj, 498, 541 

\bibitem[Kennicutt(1998b)]{Kennicutt98b} Kennicutt, R.~C., Jr.\ 1998b, \araa, 36, 189 

\bibitem[Kennicutt et al.(2003)]{Kennicutt03} Kennicutt, R.~C., Jr., et al.\ 2003, \pasp, 115, 928 

\bibitem[Kennicutt et al.(2007)]{Kennicutt07} Kennicutt, R.~C., Jr., et al.\ 2007, \apj, 671, 333 

\bibitem[Kennicutt et al.(2009)]{Kennicutt09} Kennicutt, R.~C., et al.\ 2009, \apj, 703, 1672 

\bibitem[Knapp \& Raimond(1984)]{Knapp84} Knapp, G.~R., \& Raimond, E.\ 1984, \aap, 138, 77 


\bibitem[Krumholz et al.(2009)]{Krumholz09} Krumholz, M.~R., McKee, C.~F., \& Tumlinson, J.\ 2009, \apj, 699, 850 

\bibitem[Leroy et al.(2008)]{Leroy08} Leroy, A.~K., Walter, F., Brinks, E., Bigiel, F., de Blok, W.~J.~G., Madore, B., \& Thornley, M.~D.\ 2008, \aj, 136, 2782 



\bibitem[Maraston et al.(2006)]{Maraston06} Maraston, C., Daddi, E., Renzini, A., Cimatti, A., Dickinson, M., Papovich, C., Pasquali, A., \& Pirzkal, N.\ 2006, \apj, 652, 85

\bibitem[Maraston et al.(2010)]{Maraston10} Maraston, C., Pforr, J., Renzini, A., Daddi, E., Dickinson, M., Cimatti, A., \& Tonini, C.\ 2010, \mnras, 407, 830 

\bibitem[Martin \& Kennicutt(2001)]{Martin01} Martin, C.~L., \& Kennicutt, R.~C., Jr.\ 2001, \apj, 555, 301 

\bibitem[Matthews et al.(2005)]{Matthews05} Matthews, L.~D., Gao, Y., Uson, J.~M., \& Combes, F.\ 2005, \aj, 129, 1849 


\bibitem[Neistein \& Dekel(2008)]{Neistein08} Neistein, E., \& Dekel, A.\ 2008, \mnras, 383, 615 


\bibitem[Noeske et al.(2007a)]{Noeske07a} Noeske, K.~G., et al.\ 2007, \apjl, 660, L43 

\bibitem[Noeske et al.(2007b)]{Noeske07b} Noeske, K.~G., et al.\ 2007, \apjl, 660, L47 


\bibitem[Oliver et al.(2010)]{Oliver10} Oliver, S., et al.\ 2010, \mnras, 405, 2279 

\bibitem[Oosterloo et al.(2010)]{Oosterloo10} Oosterloo, T., et al.\ 2010, \mnras, 409, 500 




\bibitem[Papadopoulos \& Pelupessy(2010)]{Papadopoulos10} Papadopoulos, P.~P., \& Pelupessy, F.~I.\ 2010, \apj, 717, 1037 

\bibitem[Papovich et al.(2010)]{Papovich10} Papovich, C., Finkelstein, S.~L., Ferguson, H.~C., Lotz, J.~M., \& Giavalisco, M.\ 2010, arXiv:1007.4554 


\bibitem[Paturel et al.(2003)]{Paturel03} Paturel, G., Petit, C., Prugniel, P., Theureau, G., Rousseau, J., Brouty, M., Dubois, P., \& Cambr{\'e}sy, L.\ 2003, \aap, 412, 45 








\bibitem[Robertson et al.(2006)]{Robertson06} Robertson, B., Bullock, J.~S., Cox, T.~J., Di Matteo, T., Hernquist, L., Springel, V., \& Yoshida, N.\ 2006, \apj, 645, 986 
\bibitem[Rodighiero et al.(2010)]{Rodighiero10} Rodighiero, G., et al.\ 2010, \aap, 518, L25 

\bibitem[Roychowdhury et al.(2009)]{Roychowdhury09} Roychowdhury, S., Chengalur, J.~N., Begum, A., \& Karachentsev, I.~D.\ 2009, \mnras, 397, 1435 

\bibitem[Ryder \& Dopita(1994)]{Ryder94} Ryder, S.~D., \& Dopita, M.~A.\ 1994, \apj, 430, 142 


\bibitem[Sage \& Welch(2006)]{Sage06} Sage, L.~J., \& Welch, G.~A.\ 2006, \apj, 644, 850 

\bibitem[Schiminovich et al.(2010)]{Schiminovich10} Schiminovich, D., et al.\ 2010, \mnras, 1288 

\bibitem[Schmidt(1959)]{Schmidt59} Schmidt, M.\ 1959, \apj, 129, 243 



\bibitem[Schuster et al.(2007)]{Schuster07} Schuster, K.~F., Kramer, C., Hitschfeld, M., Garcia-Burillo, S., \& Mookerjea, B.\ 2007, \aap, 461, 143 

\bibitem[Scoville et al.(1997)]{Scoville97} Scoville, N.~Z., Yun, M.~S., \& Bryant, P.~M.\ 1997, \apj, 484, 702 


\bibitem[S{\'e}rsic \& Pastoriza(1967)]{Sersic67} S{\'e}rsic, J.~L., \& Pastoriza, M.\ 1967, \pasp, 79, 152 

\bibitem[Shapiro et al.(2010)]{Shapiro10} Shapiro, K.~L., et al.\ 2010, \mnras, 402, 2140
 
\bibitem[Solomon et al.(1987)]{Solomon87} Solomon, P.~M., Rivolo, A.~R., Barrett, J., \& Yahil, A.\ 1987, \apj, 319, 730 

\bibitem[Springel \& Hernquist(2005)]{Springel05} Springel, V., \& Hernquist, L.\ 2005, \apjl, 622, L9 

\bibitem[Stark et al.(2009)]{Stark09} Stark, D.~P., Ellis, R.~S., Bunker, A., Bundy, K., Targett, T., Benson, A., \& Lacy, M.\ 2009, \apj, 697, 1493 

\bibitem[Shi et al.(2008)]{Shi08} Shi, Y., Rieke, G., Donley, J., Cooper, M., Willmer, C., \& Kirby, E.\ 2008, \apj, 688, 794 

\bibitem[Silk(1997)]{Silk97} Silk, J.\ 1997, \apj, 481, 703 

\bibitem[Swinbank et al.(2010)]{Swinbank10} Swinbank, A.~M., et al.\ 2010, \mnras, 405, 234 

\bibitem[Tacconi et al.(2008)]{Tacconi08} Tacconi, L.~J., et al.\ 2008, \apj, 680, 246


\bibitem[Thompson et al.(2005)]{Thompson05} Thompson, T.~A., Quataert, E., \& Murray, N.\ 2005, \apj, 630, 167 


\bibitem[Gnedin \& Kravtsov(2010)]{Gnedin10} Gnedin, N.~Y., \& Kravtsov, A.~V.\ 2010, \apj, 714, 287 

\bibitem[Graci{\'a}-Carpio et al.(2007)]{Gracia-Carpio07} Graci{\'a}-Carpio, J., Planesas, P., \& Colina, L.\ 2007, \aap, 468, L67 

\bibitem[van der Kruit \& Searle(1981)]{vanderKruit81} van der Kruit, P.~C., \& Searle, L.\ 1981, \aap, 95, 105 

\bibitem[van Driel \& van Woerden(1991)]{vanDriel91} van Driel, W., \& van Woerden, H.\ 1991, \aap, 243, 71 

\bibitem[V{\'a}zquez \& Leitherer(2005)]{Vazquez05} V{\'a}zquez, G.~A., \& Leitherer, C.\ 2005, \apj, 621, 695 

\bibitem[Walter et al.(2008)]{Walter08} Walter, F., Brinks, E., de Blok, W.~J.~G., Bigiel, F., Kennicutt, R.~C., Thornley, M.~D., \& Leroy, A.\ 2008, \aj, 136, 2563 

\bibitem[Wei et al.(2010)]{Wei10} Wei, L.~H., Vogel, S.~N., 
Kannappan, S.~J., Baker, A.~J., Stark, D.~V., 
\& Laine, S.\ 2010, \apjl, 725, L62 

\bibitem[Wong \& Blitz(2002)]{Wong02} Wong, T., \& Blitz, L.\ 2002, \apj, 569, 157 

\bibitem[Wyder et al.(2009)]{Wyder09} Wyder, T.~K., et al.\ 2009, \apj, 696, 1834 

\bibitem[Wu et al.(2005)]{Wu05} Wu, J., Evans, N.~J., II, Gao, Y., Solomon, P.~M., Shirley, Y.~L., \& Vanden Bout, P.~A.\ 2005, \apjl, 635, L173 


\bibitem[Yun et al.(1994)]{Yun94} Yun, M.~S., Scoville, N.~Z., \& Knop, R.~A.\ 1994, \apjl, 430, L109 

\bibitem[Yun \& Scoville(1995)]{Yun95} Yun, M.~S., \& Scoville, N.~Z.\ 1995, \apjl, 451, L45 

\bibitem[Zheng et al.(2007)]{Zheng07} Zheng, X.~Z., Bell, E.~F., Papovich, C., Wolf, C., Meisenheimer, K., Rix, H.-W., Rieke, G.~H., \& Somerville, R.\ 2007, \apjl, 661, L41 



\end{thebibliography}
\end{document}